\documentclass[10pt,a4paper]{article}
\usepackage[top=0.5in,left=0.6in,footskip=0.5in]{geometry}

\usepackage{amsmath,amssymb}

\usepackage{changepage}

\usepackage{textcomp,marvosym}

\usepackage{cite}

\usepackage{nameref,hyperref}

\usepackage[right]{lineno}

\usepackage[nopatch=eqnum]{microtype}
\DisableLigatures[f]{encoding = *, family = * }

\usepackage[table]{xcolor}

\usepackage{array}

\newcolumntype{+}{!{\vrule width 2pt}}

\newlength\savedwidth

\raggedright
\usepackage[aboveskip=1pt,labelfont=bf,labelsep=period,justification=raggedright,singlelinecheck=off]{caption}

\makeatletter
\renewcommand{\@biblabel}[1]{\quad#1.}
\makeatother

\usepackage{lastpage,fancyhdr,graphicx}
\usepackage{epstopdf}
\fancyheadoffset[L]{2.25in}
\fancyfootoffset[L]{2.25in}
\begin{document}
\vspace*{0.2in}

\begin{flushleft}
{\Large
\textbf{From Heterogeneity to Function: 
Enhanced Modular Dynamic Range in Hierarchical Brain Networks}
}
\newline
\\
Filipe V. Torres\textsuperscript{1,},
Mauro Copelli\textsuperscript{1},
Leonardo L. Gollo\textsuperscript{2, 3,}*
\\
\bigskip
\textbf{1} Departamento de Física, Centro de Ciência Exatas e da Natureza, Universidade Federal de Pernambuco, Recife, PE 50670-901, Brazil.
\\
\textbf{2} Instituto de Física Interdisciplinar y Sistemas Complejos, IFISC (UIB-CSIC), Palma de Mallorca, Campus Universitat de les Illes Balears, Spain.
\\
\textbf{3} Brain Networks and Modelling Laboratory, The Turner Institute for Brain and Mental Health, School of Psychological Sciences, and Monash Biomedical Imaging, Monash University, Victoria 3168, Australia.
\\
\bigskip


*\texttt{leonardo@ifisc.uib-csic.es}

\end{flushleft}


%
\section*{Abstract}
Hierarchical and modular organization is a defining feature of brain architecture, yet how it supports flexible computation and functional specialization remains largely unclear. Likewise, although dynamic range is a hallmark of efficient information processing, it has been studied predominantly as a global network property, leaving how it is distributed across modules and contributes to computation within hierarchical levels largely unexplored. Here, we show that hierarchical-modular networks naturally give rise to heterogeneous modular dynamics and firing rates, generating extended regimes of critical-like dynamics with enhanced intrinsic timescales, reflecting critical slowing down, and diverse scaling exponents, without fine-tuning when the system operates near a critical point.  Progressively longer timescales emerge at higher hierarchical levels, generating a hierarchy of intrinsic timescales, and spread of timescales within hierarchical levels arises only for networks with heterogeneity in the connectivity profile and on firing rate. We introduce the concept of modular dynamic range to characterize how individual modules respond to external inputs and show that heterogeneity enhances both global and module-specific sensitivity, with module-specific intrinsic timescales increasing with dynamic range. Under local stimulation of a single module, dynamic range is reduced in homogeneous networks but enhanced in heterogeneous networks relative to global stimulation, in which all modules receive external input. Moreover, varying the hierarchical level of the stimulated module reveals a principle of functional specialization: Selective stimulation of intermediate hierarchical levels maximizes both global and modular dynamic range. Modular dynamic range is more variable in heterogeneous networks and is enhanced in modules that do not receive direct stimulation, revealing distributed computational benefits that extend beyond the stimulated region. Together, these results show that heterogeneity in hierarchical-modular networks can extend critical-like dynamics while generating hierarchical timescales and a gradient of modular dynamic range, providing a mechanistic basis for distributed computation and functional specialization without fine-tuning to criticality.

\section*{Author summary}


{


The remarkable flexibility of the brain is supported by its hierarchical and modular architecture, yet how this structural organization enables efficient and flexible computation remains poorly understood. Using computational excitable models of hierarchical-modular networks, we show that heterogeneity in neural dynamics is not merely biological noise or variability, but a fundamental design principle. Hierarchical-modular networks can naturally generate a hierarchy of intrinsic timescales, with progressively slower dynamics emerging at higher hierarchical levels, while extending the regime of critical-like dynamics without the need for fine-tuning. We introduce the concept of modular dynamic range to quantify how individual modules respond to external inputs and show that heterogeneity enhances both local and global sensitivity to stimulation. Moreover, selective stimulation of specific hierarchical levels can maximize network-wide and module-specific responses, revealing a potential mechanism for functional specialization. These computational benefits extend beyond directly stimulated modules, demonstrating how local interactions can produce distributed enhancements in network responsiveness. Together, our findings suggest that heterogeneity is not a biological imperfection to be averaged away, but a fundamental organizational principle that supports hierarchical timescales, distributed information processing, and flexible computation.

}

\linenumbers

\section{Introduction}

The brain must support a remarkable repertoire of functions and behaviors, flexibly adapting to changing internal and external demands. A compelling theoretical framework to account for this versatility is the critical brain hypothesis, which posits that neural activity operates at, or near, a critical point of a phase transition~\cite{chialvo2010emergent, cocchi2017criticality, munoz2018colloquium}. In this regime, neural systems exhibit enhanced fluctuations, critical slowing down, and scale-free spatiotemporal correlations, conferring functional advantages such as maximized dynamic range, information capacity, and sensitivity to perturbations~\cite{beggs2008criticality,shew2013functional}. Empirical signatures consistent with criticality have been reported across species~\cite{beggs2008criticality,cocchi2017criticality,petermann2009spontaneous}, spatial and temporal scales~\cite{linkenkaer2001,palva2014correlation}, and behavioral states~\cite{priesemann2013neuronal}, supporting the view that the brain may harness the computational benefits of critical dynamics~\cite{shew2013functional,hengen2025criticality}.

A central challenge to this hypothesis is the requirement for precise fine-tuning to a critical point~\cite{rubinov2011neurobiologically,Moretti2013griffiths ,tian2022theoretical, rusch2025influence, Beggs03}. Biological systems are inherently noisy, variable, and structurally heterogeneous, rendering such fine-tuning implausible. An alternative mechanism is provided by the Griffiths phase (GP)~\cite{griffiths1969nonanalytic,vojta2006rare}, a rare-region effect in disordered systems that gives rise to extended regimes of critical-like dynamics without exact tuning~\cite{juhasz2012rare, munoz2010griffiths, Moretti2013griffiths,odor2015griffiths, vojta2006rare, odor2016critical,buendia2022broad}. In these regimes, localized islands of atypical activity dominate long-term system behavior, producing non-universal power laws, slow relaxation, and enhanced susceptibility over broad parameter ranges. GPs therefore offer a robust route for complex systems to exploit the computational benefits of criticality without requiring precise adjustment to a singular critical point.

\begin{figure}
\includegraphics[width=1.0\textwidth]{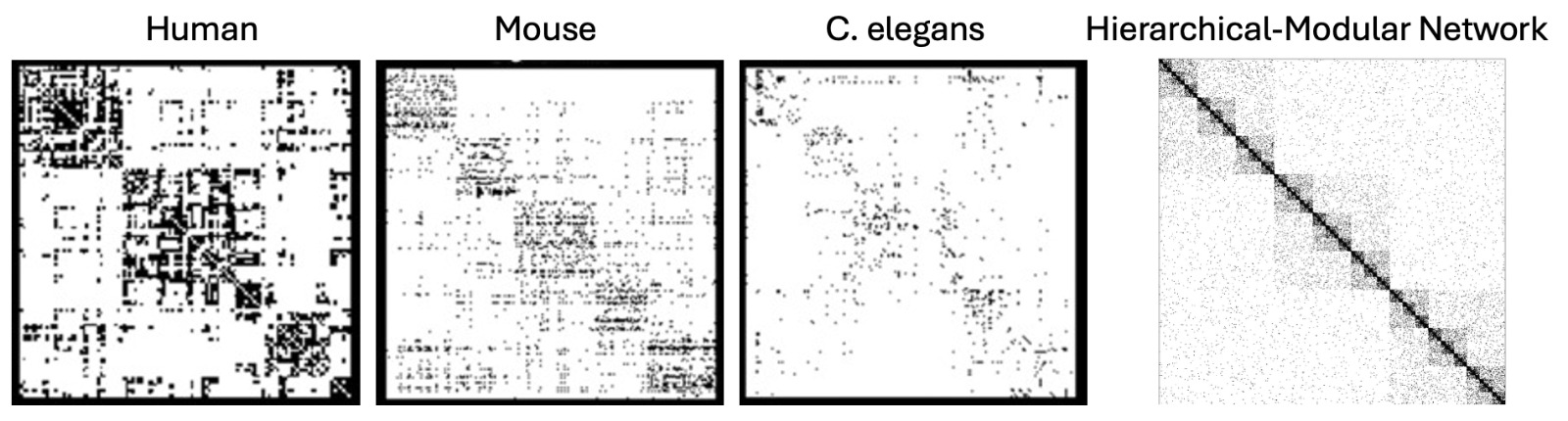}
\caption{Empirical structural connectivity matrices (connectomes) from human, mouse, and \textit{C. elegans}~\cite{zalesky2016connectome}, together with a representative synthetic example of a hierarchical-modular network (HMN) with $M_0=6$, $m=3$, $h=8$, $\alpha=2$, and $\phi=0.25$. Graph-theoretic properties are presented in Table~\ref{table1} for these empirical connectomes, and in Table~\ref{table2} for the HMN.
}
\label{figConnectomes}
\end{figure}

Theoretical studies have shown that hierarchical-modular network architectures~\cite{Kaiser2010optimal,Meunier2010Modular,kaiser2023connectomes,pena2020optimal,muller2008organization}, a hallmark of brain organization (shown in Fig.~\ref{figConnectomes}) that resembles properties of connectomes fom various species~\cite{van2016comparative}, can naturally sustain broad Griffiths-like phases~\cite{rubinov2011neurobiologically, Moretti2013griffiths, odor2016critical, wu2025optimal}, giving rise to extended regions of scale-free dynamics. In this setting, heterogeneity (acting as quenched disorder) plays a central role in shaping both the onset and extent of GP regimes. In the brain, such heterogeneity spans multiple scales, from microscopic variability in neuronal and synaptic properties to macroscopic diversity across cortical areas and intrinsic neural timescales (INTs). Crucially, this variability is a structured feature that may enable critical-like dynamics to emerge under biologically realistic conditions. A particularly relevant example is the hierarchy of intrinsic timescales across cortical regions~\cite{kiebel2008hierarchy,murray2014hierarchy,gollo2015dwelling, chaudhuri2015large,muller2023spatial,harris2024tracking}, which reflects mesoscale heterogeneity tied to network architecture and shapes the temporal organization of neural activity across scales~\cite{raut2020hierarchical,cavanagh2020diversity,wu2024dendrites}. Typically estimated from the decay of the autocorrelation function~\cite{golesorkhi2021temporal,wolff2022intrinsic, golesorkhi2021brain,wu2025mapping,wu2025criticality}, INTs have been linked to criticality via critical slowing down~\cite{wu2025mapping,wu2025criticality}, whereby temporal correlations increase in the vicinity of a critical region~\cite{scheffer2009early, dakos2008slowing,liu2024early}.

\begin{table}
    \centering
    \renewcommand{\arraystretch}{1.2}
    \begin{tabular}{c||c|c|c|c|c}
Species & Density & Modularity & Efficiency & Clustering & Nodes\\
\hline
Human & 22\% & 0.38 & 0.56 & 0.58 & 116\\
Mouse & 6.8\% & 0.43 & 0.39 & 0.29 & 213\\
\textit{C. elegans} & 3.8\% & 0.41 & 0.27 & 0.17 & 279\\
\end{tabular}
    \\
    \vspace{1mm}
    \caption{Graph-theoretic properties of empirical connectomes shown in Fig.~\ref{figConnectomes}.
    Measures for HMNs are presented in Table~\ref{table2}.
    }
    \label{table1}
\end{table}

One of the most striking dynamical properties shared by systems both at criticality and in nearby GP-like regimes is an enhanced dynamic range~\cite{Kinouchi06a,Moretti2013griffiths}, which quantifies the capacity to generate discriminable responses to varying external inputs. In a notable example of theory–experiment interplay in computational neuroscience, maximization of dynamic range was first predicted theoretically~\cite{Kinouchi06a} and subsequently confirmed experimentally~\cite{shew2009neuronal}. Multiple structural and dynamical factors have been shown to amplify and modulate dynamic range, including network size~\cite{gollo2012statistical, furtado2006response}, dimensionality~\cite{Assis08,ribeiro2008deterministic}, tree-like branching~\cite{copelli2007excitable,gollo2009active}, degree distribution~\cite{wu2007excitable, Larremore11a}, motif connectivity~\cite{copelli2007excitable}, and excitability~\cite{gollo2012signal,publio2012dynamic}. Moreover, heterogeneity can alter the nature of the phase transition itself, further enhancing both dynamic range~\cite{gollo2016diversity} and the dynamic-range-to-noise ratio~\cite{gollo2017coexistence}. Despite this growing body of work, the global effects of heterogeneity and dimensionality on broad Griffiths-like dynamics in hierarchical-modular networks, as well as the local impact of selective stimulation across modules and hierarchical levels, remain poorly understood.

Here, we systematically investigate how the dynamics of broad scaling regimes in hierarchical-modular networks that mirror brain architecture depend on local heterogeneity in the dynamics of modules and global structural features of the network. We begin by introducing these networks and their key properties, including network dimensionality, which shapes the underlying dynamics and governs heterogeneity in firing rates. This heterogeneity, in turn, extends the range of coupling strengths that give rise to power-law behavior and enhances long-range temporal correlations, consistent with critical slowing down. We show that intrinsic neural timescales (INT) become progressively longer at higher hierarchical levels. Finally, we characterize the dynamic range of the full network and its subcomponents under selective stimulation, identifying conditions that optimize performance at local modular level and for the overall system. By integrating statistical physics with systems neuroscience, our work establishes a mechanistic framework for how structural disorder of hierarchical-modular networks can promote heterogeneous and robust, critical-like dynamics. This framework not only explains the emergence of hierarchical timescales and enhanced dynamic range, but also clarifies how functional diversity can arise without strict fine-tuning. Our findings advance the critical-brain hypothesis by demonstrating how the brain can sustainably harness the benefits of criticality through its inherent and ubiquitous disordered structure.

    




\section{Results}

\begin{figure}
\includegraphics[width=0.99\textwidth]{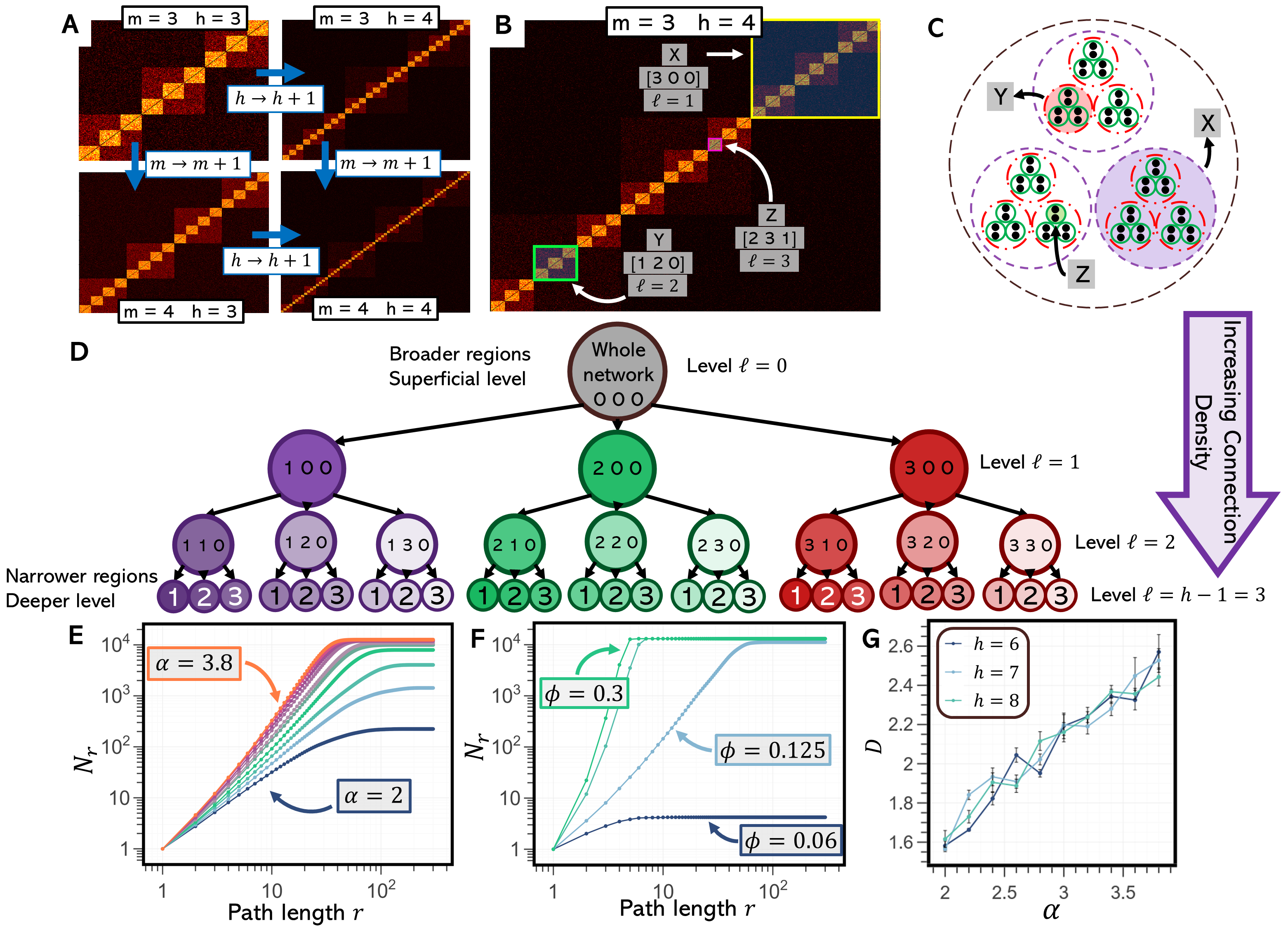}
\caption{
\textbf{A}: Average of 20 adjacency matrices for varying $m$ and $h$. In each matrix, $m$ squared blocks repeat at every sub-level $\ell$, up to $\ell = h$. Each entry of the matrix represents a connection between two neurons.
\textbf{B}: Example network with $m=3$ and $h=4$. The numbers label nested modules, and the location of a module is given by reading these labels from the outermost level inward, e.g., $Z$: $[2,3,1]$.
\textbf{C}: Schematic of the same modular structure as in \textbf{B}, with $M_0=2$; physically closer modules are more densely connected.
\textbf{D}: Hierarchical representation of the network, where the hierarchy corresponds to the modular coordinates with trailing zeros.
\textbf{E}, \textbf{F}: Number of nodes reached, $N_r$, as a function of path length $r$, for fixed $\phi = 2^{-3} = 0.125$ with varying $\alpha$ (\textbf{E}), and for fixed $\alpha = 3$ with varying $\phi$ (\textbf{F}).
In \textbf{E}, $\alpha$ controls the slope of $N_r$ on the log--log plot, which defines the network dimension $D$.
In \textbf{F}, $N_r$ follows a power law in $r$ only for $\phi = 0.125$.
\textbf{G}: Network dimension, computed with Dijkstra's algorithm.
Unless otherwise stated, network parameters are $M_0=6$, $m=3$, and $h=8$.
}
\label{fig:network}
\end{figure}

We investigate the dynamical properties that arise in hierarchical-modular networks (Figs.~\ref{figConnectomes} and~\ref{fig:network}), proposed by Moretti and Muñoz~\cite{Moretti2013griffiths}, and their role in facilitating efficient computation. The inherent organization of these networks offers a robust and versatile framework for generating a broad spectrum of biologically relevant dynamical behaviors without requiring precise parameter tuning. We introduce the fundamental structural (Tables~\ref{table1} and~\ref{table2}) and dynamical characteristics of these networks, emphasizing the emergence of heterogeneous dynamics, and explore the implications of this heterogeneity of dynamics across modules. We focus on how this heterogeneity shapes the  dynamic repertoire of the network and its components, contributes to a rich diversity of INT, and improves the dynamic range of the network. 

\subsection{Hierarchical-modular networks (HMN)} 
\label{subsec:network_def}

We study the dynamics of hierarchical-modular networks in which each network has $N$ nodes with $m$ modules, $h$ hierarchical levels, and $M_0$ nodes in the innermost modules (Fig.~\ref{fig:network}~\textbf{A-D}), such that $N=M_0 \times m^{h-1}$. Connectivity depends on the hierarchical level $\ell=\{0,1,2,\dots,h-1\}$, with the connection density $\phi_\ell$ within each module given by $\phi_\ell=\alpha \cdot \phi^{h-\ell}$, where $\alpha$ and $\phi$ are control parameters (Fig.~\ref{fig:network}~\textbf{E, F}).

The parameter $\alpha$ determines the finite network dimension $D$ (Fig.~\ref{fig:network}~\textbf{G}) and is related to the network's mean degree~\cite{odor2015griffiths}. The network dimension is a key metric that characterizes the scaling of the ``hyper-sphere'' occupied by the network nodes~\cite{emmerich2013complex, daqing2011dimension}, i.e., how the number of reached nodes $N_r$ increases with path length $r$: $N_r \sim r^D$. Low-$D$ networks rapidly saturate their node count, whereas infinite-dimensional networks exhibit growth faster than any power law. Networks with $\phi > 2^{-m}$ are considered infinite-dimensional regardless of the value of $\alpha$~\cite{odor2015griffiths, Moretti2013griffiths}. Throughout this work, we contrast these networks with their finite-dimensional counterparts to investigate the impact of structural heterogeneity on network dynamics.

The parameter $\phi$ directly controls the network's sparsity, with lower values producing sparser connectivity and higher values producing denser connectivity. Since $\phi < 1$, the connection density decreases with increasing hierarchical level, giving rise to progressively coarser modular organization, as illustrated by the adjacency matrices in Fig.~\ref{fig:network}~\textbf{A, B}.

The HMN topology naturally defines modules according to their location in the adjacency matrix, with each module corresponding to a specific set of neurons. To simplify their representation, we introduce a modular coordinate to identify the module to which each neuron belongs, complementing the traditional neuron indexing from $1$ to $N$. This coordinate consists of a sequence of $h-1$ numbers ranging from $1$ to $m$, where each digit specifies the module index at its corresponding hierarchical level. This representation allows us to naturally assign and analyze both neuronal activity and external stimuli at the module level. See Section~\ref{subsecCoordinates} for further details.

\subsection{Sparsity enhances dynamical heterogeneity in hierarchical-modular networks} 

The neuronal dynamics is simulated using the susceptible-infected-susceptible (SIS) model, in which each neuron is represented as a two-state cellular automaton: resting ($s_i=0$) or excited ($s_i=1$). A neuron becomes excited either through an external Poisson stimulus with rate $r$, occurring with probability $\eta(r)=1-e^{-r}$ at each time step ($\delta_t=1$ms), or through excitation from an active neighboring neuron with probability $p$. The resulting dynamics generate a time-dependent density of active sites, $\rho(t)$, whose temporal average corresponds to the network activity ($F$).

The connectivity matrix $\textbf{C}$ is the adjacency matrix $\textbf{A}$ multiplied by propagation probability $p$ (i.e., $\textbf{C} = p \textbf{A}$). The key control parameter governing the dynamics is $\lambda$, defined as the largest eigenvalue of $\textbf{C}$~\cite{Larremore11a}.

\begin{figure}
    \centering
    \includegraphics[width=1.0\textwidth]{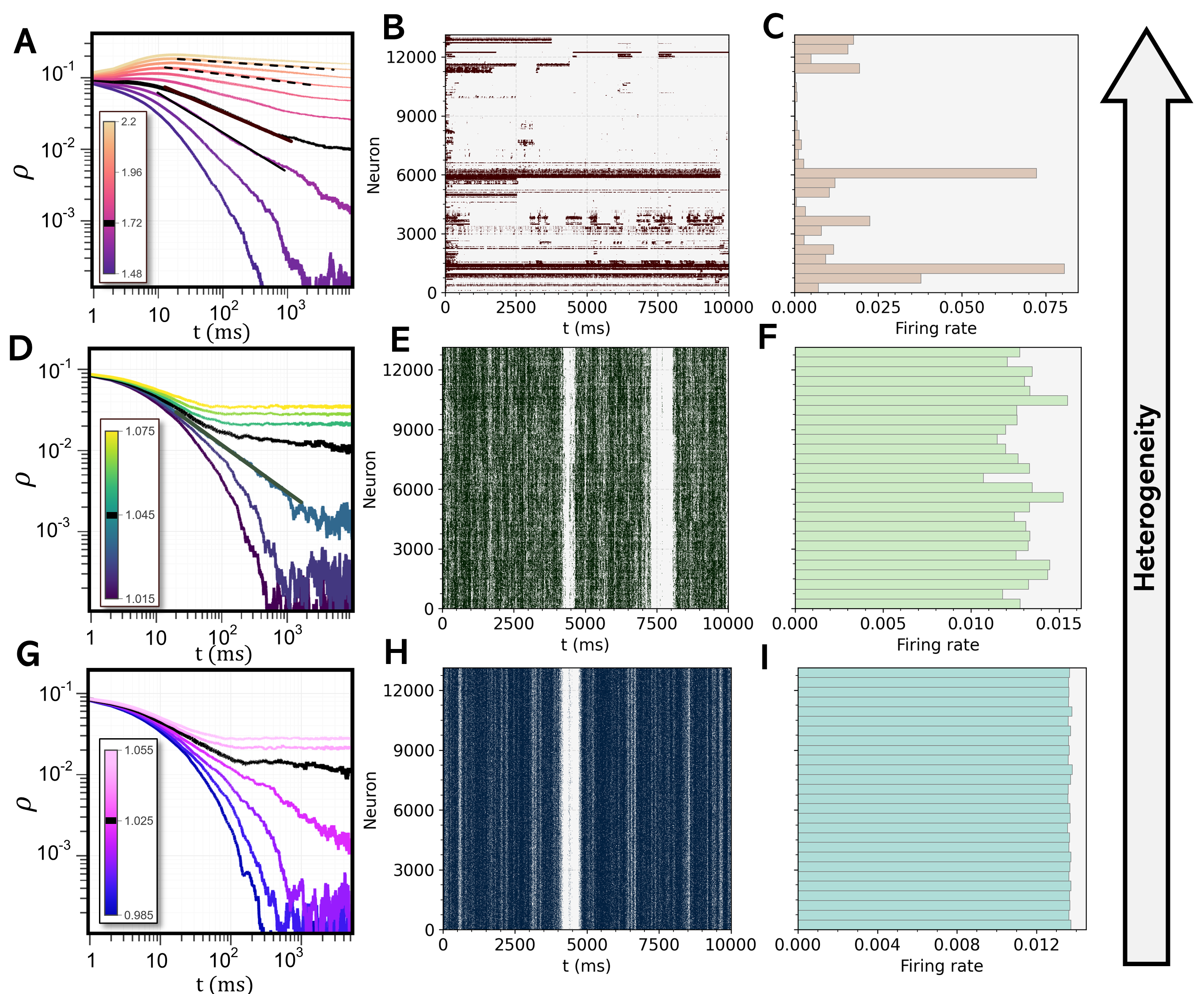}
    \caption{\textbf{Raster plots and time series comparison of heterogeneity of each network, from homogeneous networks to highly heterogeneous ones.} The first row (\textbf{A}, \textbf{B}, \textbf{C}) illustrates a network, exhibiting a continuum of power-law exponents and pronounced heterogeneity in local firing rates. The second row (\textbf{D}, \textbf{E}, \textbf{F}) corresponds to mildly heterogeneous models, which display a single power-law decay and substantially weaker heterogeneity, while the third row (\textbf{G}, \textbf{H}, \textbf{I}) conveys the completely homogeneous case, with single power laws and no heterogeneity in activity. Panels on the left, i.e., \textbf{A}, \textbf{D} and \textbf{G} show time series of the density of active sites, defined as the fraction of active nodes in the network under a weak external stimulus of $r = 10^{-6}\text{ms}^{-1}$.  For the $\lambda$ value corresponding to the black curves, one representative realization (out of the 100 averaged trials) is selected, and the corresponding raster plots are shown in \textbf{B} (with $p=0.617$), \textbf{E} (with $p=0.138$) and \textbf{H} (with $p=0.0051$). 
    Panels \textbf{C}, \textbf{F} and \textbf{I} present the distributions of local firing rates across network nodes. All of those were run with networks of $M_0=6$ nodes on the innermost module, $m=3$ sub-modules, $h=8$ hierarchical levels and $\alpha =2$, while $\phi$ differs: On the first row, $\phi=0.125$,  on the second, $\phi=0.25$, and on the third row $\phi=0.5$. All raster plots were performed with  $\lambda$  such that the $\rho \simeq 0.013$ in all cases.}
    \label{figtime series+rasters}
\end{figure}

In the absence of external input, the model exhibits a continuous phase transition as $\lambda$ 
is varied~\cite{Marro99}. For weak coupling, activity cannot be sustained and eventually dies out, whereas sufficiently strong coupling gives rise to persistent self-sustained activity. These distinct dynamical regimes are illustrated in panel \textbf{A} of Fig.~\ref{figtime series+rasters} for different values of the coupling strength. The raster plots and firing-rate distributions shown in panels \textbf{B} and \textbf{C} reveal a pronounced spatial heterogeneity in neuronal activity. While some neurons remain nearly inactive throughout the simulation, others exhibit consistently high firing rates, indicating the emergence of localized regions, islands of the network, with markedly different activity levels. This dynamical heterogeneity becomes progressively weaker as the hierarchical modularity parameter $\phi$ increases (Fig.~\ref{figtime series+rasters}\textbf{D}--\textbf{I}), leading to a more homogeneous distribution of neuronal activity. Networks with relatively homogeneous dynamics exhibit power-law temporal fluctuations in $\rho(t)$ characterized by a well-defined scaling exponent. In contrast, heterogeneous hierarchical-modular networks also display power-law fluctuations, but with varying scaling exponents Fig.~\ref{figtime series+rasters}~\textbf{A}, reflecting the broader spectrum of dynamical behaviors generated by the heterogeneity in neuronal activity.

While all other parameters remain fixed, the HMN in Fig.~\ref{figtime series+rasters}~\textbf{A},~\textbf{D}, and~\textbf{G} show a strong dependence on 
$\phi$, yielding substantially different graph-theoretic properties (see Section~\ref{subsecGT} for details). Table~\ref{table2} compares the values obtained for the networks whose dynamics are shown in Fig.~\ref{figtime series+rasters}. It shows that heterogeneous HMNs are more sparse, have greater modularity, and exhibit lower efficiency.
A major driver of nodal-activity heterogeneity is the coefficient of variation ($CV$) of both the in-degree distribution 
($CV_{in-degree}$) and the convergence index ($CV_{\kappa}$). 
For $\phi=0.125$, 0.25, and 0.5, the Spearman 
correlation between mean firing rate and in-degree is
$r=0.46$, $r=0.85$ and $r=0.59$, respectively, with all correlations remaining significant after Bonferroni correction ($p\text{--value} < 10^{-6}$).

\begin{table}
    \centering
    \renewcommand{\arraystretch}{1.2}
    \begin{tabular}{c||c|c|c}
                & $\phi=0.125$ &  $\phi=0.25$ &  $\phi=0.5$ \\
 \hline
      Density     & 0.014\% &  0.06\% & 1.5\% \\
     Modularity  &  0.97&    0.81 &  0.32\\
     Efficiency   & 0.0013     & 0.18     & 0.5\\
     Nodes   & 13122 & 13122 & 13122 \\
     $\langle$in-degree$\rangle$ &  1.9 & 7.7 & 198 \\
      $\mathrm{std}(\text{in-degree})$
 &  1.2 & 2.5 & 13.5 \\
      $CV_{in-degree}$ &   0.67 & 0.33 & 0.068 \\
      $\langle \kappa \rangle$ &   1.1     &  1  &   1 \\
    $CV_{\kappa}$ &   0.67 & 0.35 & 0.069 \\
    
    \end{tabular}
        \vspace{1mm}

    \caption{Graph-theoretic measures of the HMN with parameters $M_0=6$, $m=3$, $h=8$, and $\alpha=2$, corresponding to the networks shown in Fig.~\ref{figtime series+rasters}.
    The convergence index $\kappa$ is defined as the product of the in-degree and the propagation probability $p$ (used in the middle-column panels of Fig.~\ref{figtime series+rasters}). $CV$ denotes the coefficient of variation (defined as the ratio of the standard deviation to the mean), and $\mathrm{std}$ denotes the standard deviation.
    }
    \label{table2}
\end{table}

\subsection
{Heterogeneity stretches criticality and breaks the mean-field approximation }

Random networks are effectively infinite-dimensional and exhibit a continuous second-order phase transition~\cite{Kinouchi06a, Larremore11a}, with $F_0$, the mean activity in the absence of external stimulation, serving as the order parameter and $\lambda$ as the control parameter. This mean-field description of the directed percolation phase transition is exact only in the limit of homogeneous, infinite-dimensional networks~\cite{Moretti2013griffiths}. In contrast,  dynamical heterogeneity of hierarchical-modular networks can fundamentally alter the critical behavior. This is exacerbated in finite-dimensional systems but can also be relevant in the dynamics of hierarchical-modular networks  with infinite dimension. Different dynamical properties can reflect the distance to criticality~\cite{harris2024tracking,scheffer2009early}, including the autocorrelation function, as recently chararcterized in HMNs~\cite{pena2020optimal}.

\begin{figure}
    \centering
    \includegraphics[width=0.90\textwidth]{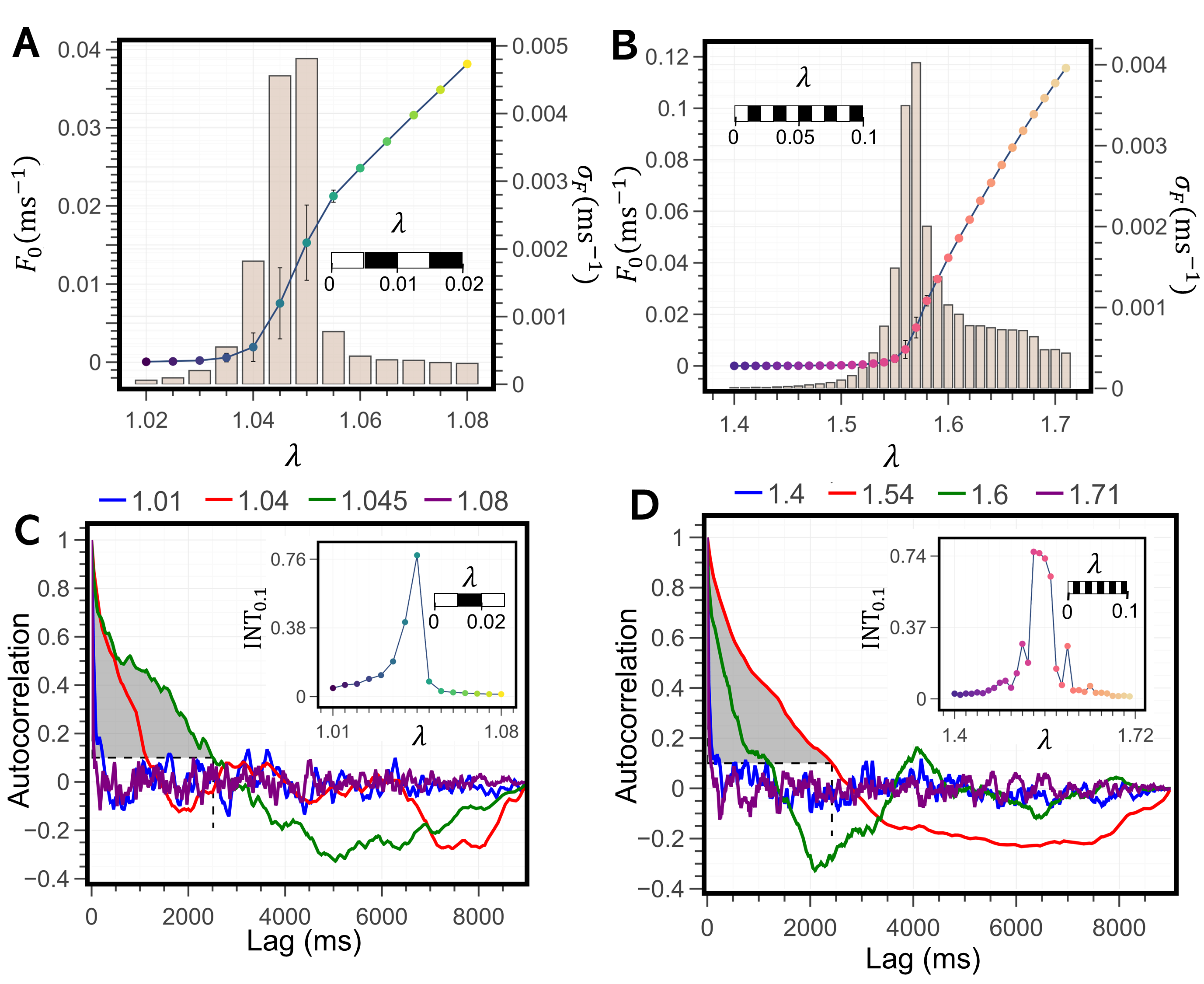}
    \caption{\textbf{Dynamical heterogeneity stretches the signatures of criticality.}
Comparison between infinite- and finite-dimensional hierarchical-modular networks. Both networks share the same modular architecture ($M_0=6$, $m=3$, $h=8$, $\alpha=2$), differing only in the parameter $\phi$: $\phi=0.25$ for the infinite-dimensional network (\textbf{A,C}) and $\phi=0.125$ for the finite-dimensional network (\textbf{B,D}) (see Fig.~\ref{fig:network}\textbf{F}). \textbf{A,B}: spontaneous activity $F_0$, measured as the average density of active sites over $T=10^4~\text{ms}$ across 50 realizations; error bars denote the standard deviation. \textbf{C,D}: temporal autocorrelation functions computed under a weak external stimulus ($r=10^{-6}\text{ms}^{-1}$). The shaded region illustrates the definition of the intrinsic neural timescale ($\text{INT}$), computed as the area under the autocorrelation curve. Insets show $\text{INT}$ as a function of the coupling strength.}
    \label{fig:whole_activity_response_INT}
\end{figure}

The dynamical effects of finite dimensionality in HMNs is illustrated in Fig.~\ref{fig:whole_activity_response_INT}\textbf{A,B}. In the infinite-dimensional network (Fig.~\ref{fig:whole_activity_response_INT}\textbf{A}), the transition between the absorbing and active phases is sharply localized around a single critical coupling. By comparison, the finite-dimensional network (Fig.~\ref{fig:whole_activity_response_INT}\textbf{B}) displays a broad region of enhanced fluctuations, reflected by the stretched peak in the standard deviation of the spontaneous activity, $\sigma_{F_0}$. These large fluctuations coincide with the extended parameter region in which power-law activity is observed, a hallmark of Griffiths-like dynamics.

Another characteristic signature of criticality is \textit{critical slowing down}, whereby perturbations decay increasingly slowly as the system approaches the critical regime. To quantify this phenomenon, we computed the temporal autocorrelation function (ACF) of the activity density $\rho(t)$ and measured the associated intrinsic neural timescale ($\text{INT}$), defined as the area under the ACF from lag $0$ to the first lag $k$ satisfying $ACF$(k)=0.1. Consistent with previous studies~\cite{wu2025mapping,wu2025criticality}, infinite-dimensional networks exhibit a sharp peak in $\text{INT}$ precisely at the critical point (Fig.~\ref{fig:whole_activity_response_INT}\textbf{C}). In contrast, finite-dimensional networks display a broad plateau of elevated $\text{INT}$ values extending across a wide range of coupling strengths (Fig.~\ref{fig:whole_activity_response_INT}\textbf{D}). Rather than peaking sharply, $\text{INT}$ decreases gradually beyond its maximum, indicating that critical slowing down persists throughout an extended parameter region. Together with the broadened fluctuations in spontaneous activity (Fig.~\ref{fig:whole_activity_response_INT}\textbf{B}) and the extended power-law regime in $\rho(t)$ (Fig.~\ref{figtime series+rasters}), these results demonstrate that dynamical heterogeneity of HMN stretches the signatures of criticality into a broad Griffiths-like phase.

\begin{figure}
    \centering
    \includegraphics[width=0.95\linewidth]{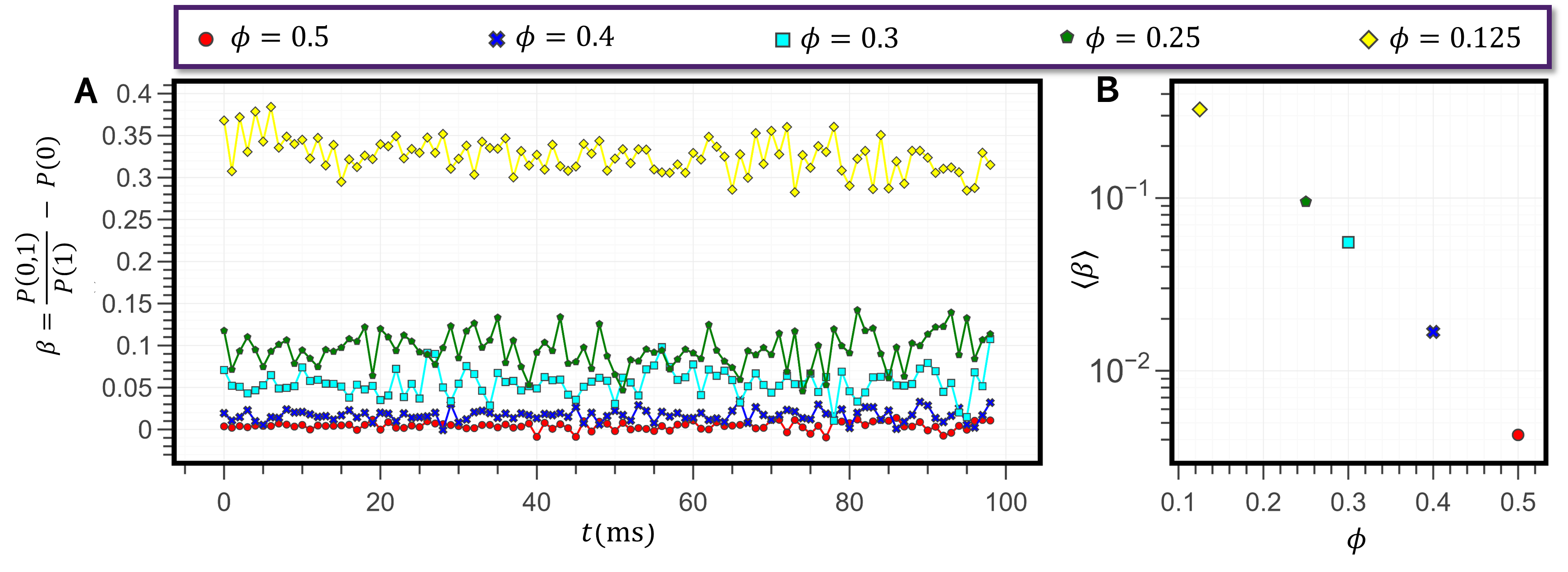}
    \caption{\textbf{Breakdown of the mean-field prediction of the critical point.} \textbf{A}: Temporal evolution of the quantity $\frac{P(0,1)}{P(0)}-P(1)$, which measures the deviation from the mean-field assumption that neighboring neuronal states are statistically independent. This quantity is therefore expected to vanish when the analytical prediction $\lambda=1$ accurately describes the network's critical point. As the network becomes increasingly heterogeneous (decreasing $\phi$), $\beta$ departs from zero, demonstrating the breakdown of the mean-field approximation~\cite{Larremore11a} and explaining the discrepancy between the predicted and observed critical coupling. \textbf{B}: Time-averaged value of $\beta$. Network parameters are: $M_0=6$, $m=3$, $h=8$, and $\alpha=2$.
    }
    \label{fig:independence_S3_fig}
\end{figure}

One of the key assumptions underlying the mean-field prediction of the critical point, $\lambda_c=1$, is that the states of neighboring neurons are statistically independent, such that
$P(0,1)\simeq P(1)P(0)$, where $P(0,1)$ denotes the probability that a presynaptic neuron is active while its postsynaptic neighbor is inactive. This mean-field approximation is valid when correlations between neighboring neurons are sufficiently weak. However, the structural heterogeneity of hierarchical-modular networks induces strong local correlations, violating this assumption and causing the true critical point to deviate from the mean-field prediction. To quantify this deviation, we define
$\beta=\frac{P(0,1)}{P(0)}-P(1)$,
which measures the breakdown of the independence assumption. As shown in Fig.~\ref{fig:independence_S3_fig}, $\beta$ increasingly departs from zero as network heterogeneity grows, explaining why the mean-field approximation becomes progressively less accurate in finite-dimensional hierarchical-modular networks.

\subsection
{A hierarchy of timescales emerges in
hierarchical-modular networks}
\label{subsecmodular_time series}

Whole-network measures provide a useful characterization of the critical behavior of hierarchical-modular networks and have helped explain several aspects of brain dynamics~\cite{Kinouchi06a,litwin2012slow,odor2015griffiths,wu2026structure}. However, they represent spatial averages over the entire network and therefore obscure the variability arising from individual modules. This limitation becomes particularly important in strongly heterogeneous networks, where disorder and rare-region effects are expected to produce substantial local differences in the dynamics. To capture these effects, we exploit the modular and hierarchical organization of the network by analyzing the activity of  modules independently. Instead of a single time series for the global activity density $\rho(t)$, we compute one time series for each module at every hierarchical level, following the procedure described in Section~\ref{subsec:network_def}. Each modular time series is then analyzed using the same procedure employed for the whole network (Fig.~\ref{fig:whole_activity_response_INT}), allowing us to compute the corresponding mean activity $F_0$, autocorrelation function, and intrinsic neural timescale (INT). This modular analysis enables us to investigate how signatures of criticality emerge locally within different regions of the network.

\begin{figure}
    \centering
    \includegraphics[width=1.0\textwidth]{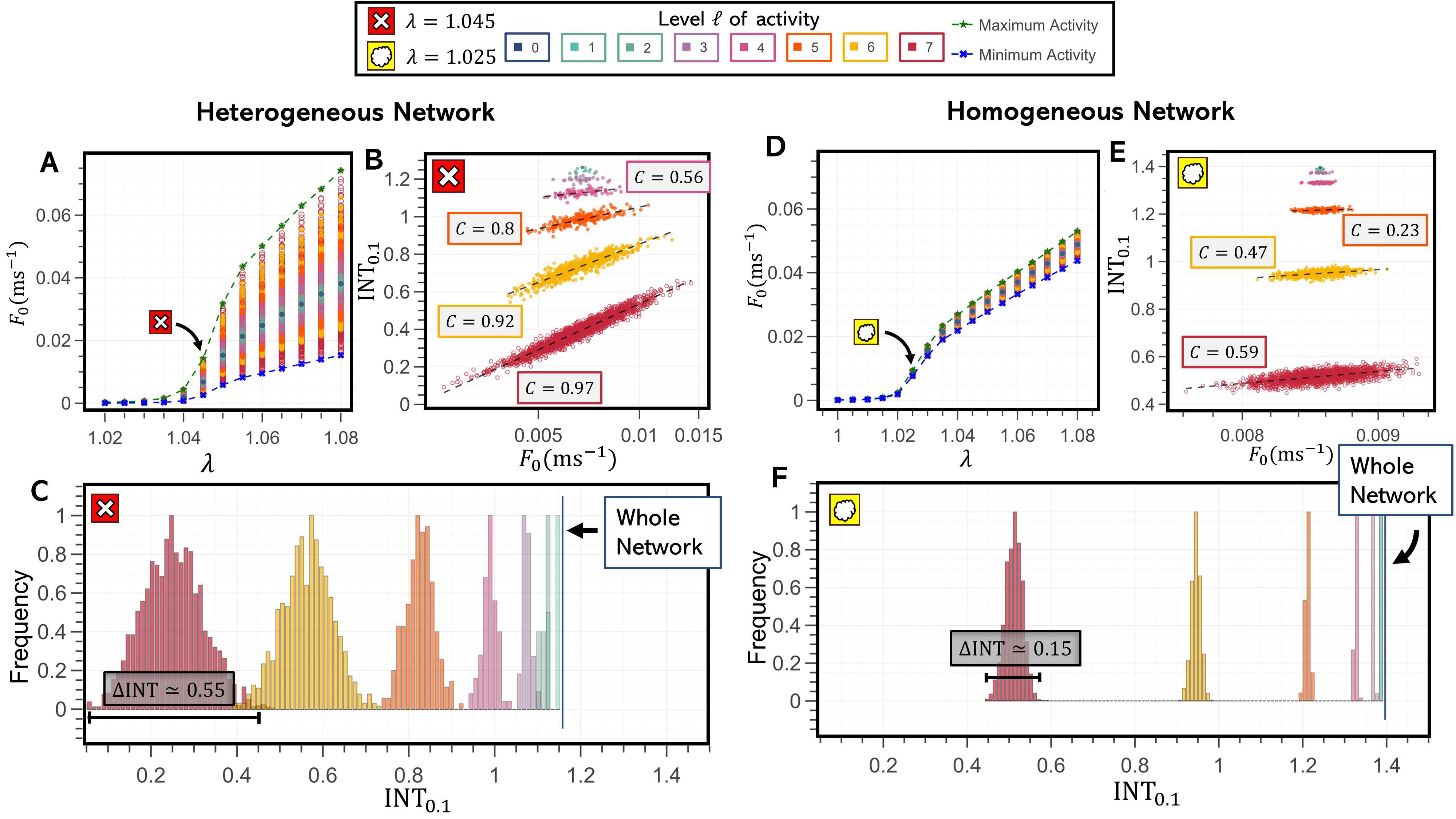}
    \caption{\textbf{Modular mean activity and local critical slowing down measures showing spread of criticality through modules.} This figure shows a comparison between mean activity and INT in heterogeneous networks ($\phi=0.25$,  \textbf{A-C}) and homogeneous ones ($\phi=0.5$, \textbf{D-F}). On \textbf{A} and \textbf{D}, mean activities $F_0$ display modules with vastly different activities between the most active one (represented by the dashed green line) and the least active one (through a dashed blue line). We then calculate $\text{INT}$ for the time series on each module on the network on the specific coupling for which global $\text{INT}$ is maximized. These $\text{INT}$ values are then plotted against $F_0$ on \textbf{B} and \textbf{C}, while the Pearson correlation $C$ is calculated for each hierarchical level and displayed whenever its correspondent p-value $p<0.05$. A histogram of frequencies is also plotted in \textbf{C} and \textbf{F}, highlighting the increased width of INT in the heterogeneous network.
     Network parameters are: $M_0=6$, $m=3$, $h=8$, and $\alpha=2$.
    }
    \label{figmod_activity}
\end{figure}

Our results reveal that  heterogeneity, controlled by $\phi$, strongly increases the diversity of module dynamics. As shown in Fig.~\ref{figmod_activity}\textbf{A}, the mean activity of individual modules in heterogeneous networks spans a broad range of values, with some overlap between neighboring hierarchical levels, in sharp contrast to the much narrower distribution observed in homogeneous networks (Fig.~\ref{figmod_activity}\textbf{D}). This broad dispersion persists beyond the global transition, indicating that different modules can simultaneously occupy distinct dynamical regimes. Rather than undergoing a collective phase transition, the disordered connectivity promotes local dynamical heterogeneity, allowing subcritical modules to coexist with others that already exhibit self-sustained activity across an extended range of coupling strengths.

This heterogeneous behavior is even more evident when examining the intrinsic neural timescales. The histograms in Fig.~\ref{figmod_activity}\textbf{C,F} show that
INT generally grows for higher hierarchical levels (lower values of $\ell$),
and that the distribution of INT values is substantially broader in heterogeneous networks than in homogeneous ones across all hierarchical levels. The effect is strongest in the innermost modules ($\ell=h-1=7$), where the range of INT values is approximately three times larger than in the homogeneous network. These results demonstrate that critical slowing down is itself spatially heterogeneous, providing direct evidence that criticality is distributed across modules rather than occurring uniformly throughout the network.

A direct consequence of this modular heterogeneity is that, even for coupling strengths near the onset of the globally active phase (e.g., $\lambda\approx1.04$ in Fig.~\ref{figmod_activity}\textbf{A}), modules at the same hierarchical level can exhibit markedly different dynamical states. While some innermost modules remain predominantly subcritical, others already display self-sustained activity. As shown in Fig.~\ref{figmod_activity}\textbf{B}, these differences are strongly associated with the corresponding INT values: modules with higher activity consistently exhibit longer intrinsic timescales. This relationship is particularly pronounced in heterogeneous networks, whereas homogeneous networks exhibit  weaker correlations together with much narrower distributions of both activity and INT (compare Fig.~\ref{figmod_activity}\textbf{B} and \textbf{E}).
Similarly to its central role in determining the network dynamics (Fig.~\ref{fig:independence_S3_fig}), $\phi$ also governs the distribution of dynamical states across modules. These findings demonstrate that decreasing $\phi$ gives rise to a hierarchy of local dynamical states, with individual modules approaching criticality (transition to the self-sustained state) at different coupling strengths.

\subsection{Heterogeneity-driven recruitment enhances stimulus discrimination}

\begin{figure}
    \centering
    \includegraphics[width=0.8\linewidth]{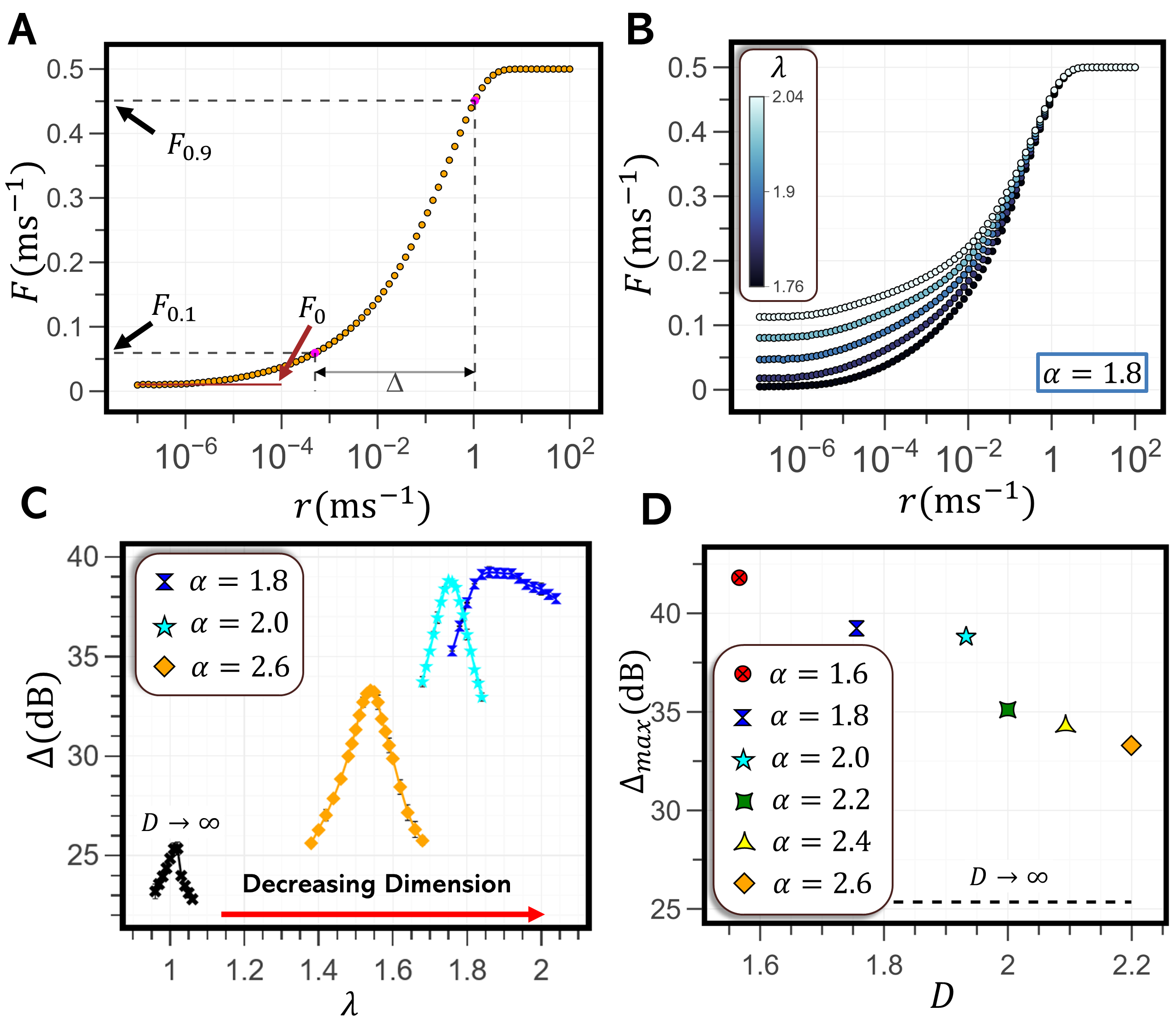}
    \caption{\textbf{Network response to stimuli, dynamic range and its dependence with topological dimension.} All results were obtained through a network of $M_0=4$, $m=2$, $h=13$ and $\phi=0.25$ while varying $\alpha$ (therefore, the topological dimension) for the finite dimensional case, and $\phi = 0.5$ for the infinite dimensional (black data on \textbf{C} and \textbf{D}), and 60 realizations were run for each
  $r$ and $\lambda$. On \textbf{A}, it is shown the response curve and its correspondent dynamic range on a network of $\alpha=2.6$. The dotted lines correspond to the coordinates of $(r_{0.1}, F_{0.1})$ and $(r_{0.9}, F_{0.9})$ obtained and used to estimate $\Delta = 33.29 \pm 0.03 \ \text{dB}$. This method is then applied for all response curves varying $\lambda$ shown on \textbf{B} for this network of $\alpha=1.8$. For comparative studies on topological dimension, by varying $\alpha$, we can repeat this process, registering the maximum dynamic range obtained, generating \textbf{C}, that shows the different values for the $\Delta$ vs. $\lambda$, and \textbf{D}, with its maximum $\Delta _\text{max}$ vs. $D$, including the black curve, of infinite topological dimension. }
    \label{figglobal_response_dyn}
\end{figure}

The brain continuously operates under a barrage of sensory inputs, making the encoding of external signals a defining feature of neural computation. To assess this capability, we extend our analysis beyond spontaneous dynamics and fixed-input simulations by systematically varying the stimulus rate $r$. The resulting relationship between the average firing rate $F$ and the stimulus rate $r$ defines the \textit{response curve} (Fig.~\ref{figglobal_response_dyn}\textbf{A}), a sigmoidal input--output function that quantifies the network's dynamic range, $\Delta$, namely the range of stimulus intensities that can be reliably discriminated by neuronal activity~\cite{Kinouchi06a}. This input discrimination is studied under two scenarios: (i) when the entire network is externally driven, and (ii) when stimulation is confined to specific modules and sub-networks of higher hierarchical levels. The systematic examination of localized external driving helps shed light on the interplay between network structure and dynamic functional specialization that emerges from heterogeneous dynamics in HMN.  

\begin{figure}
    \centering
    \includegraphics[width=0.6\linewidth]{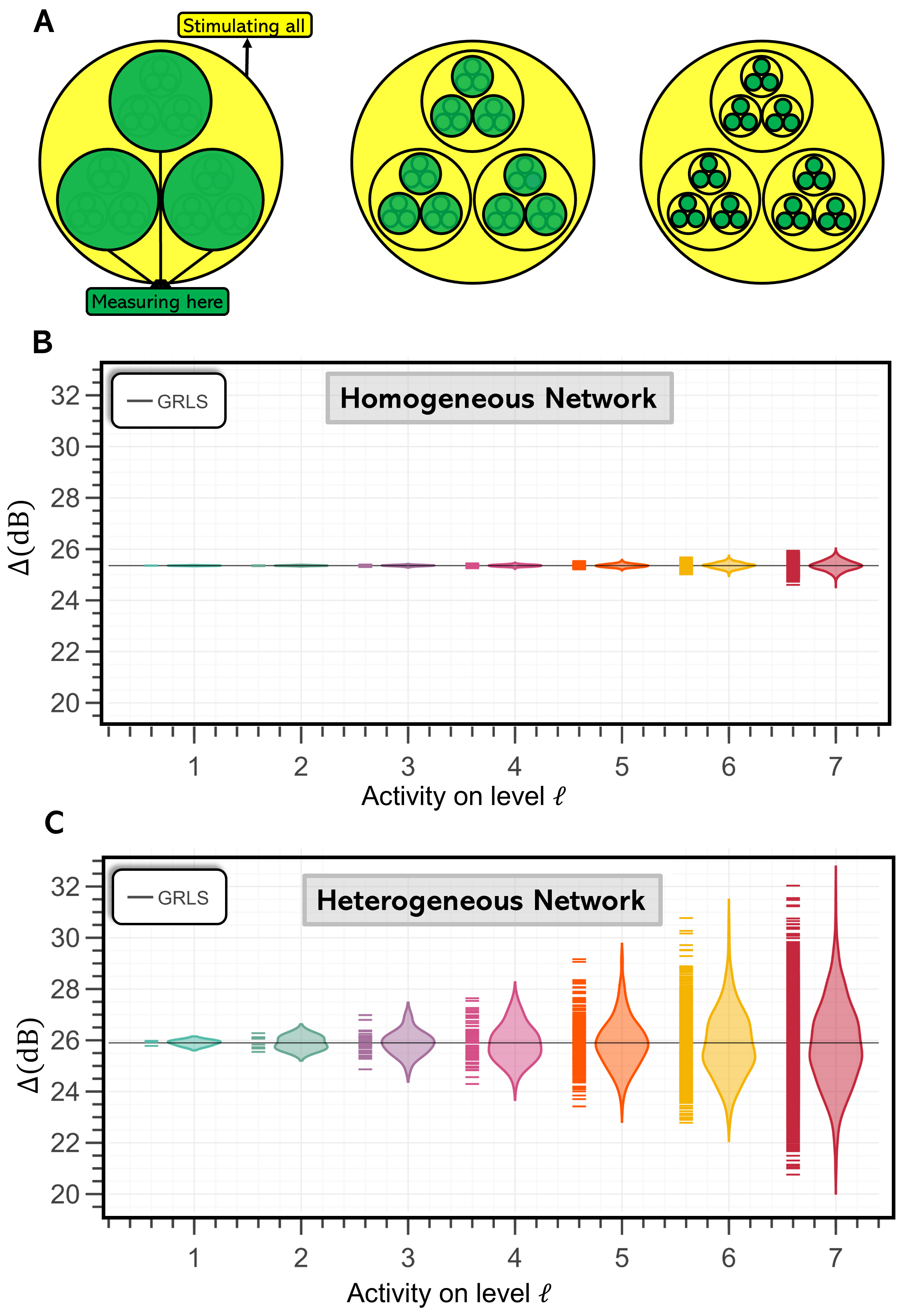}
    \caption{\textbf{Local response across modules and hierarchical levels to a global stimulus (i.e., the whole network).} Panel \textbf{A} Schematic representations of what level the measure is taken from, highlighting that the stimulation is always on all of them. By grouping the response curves as to the hierarchical level of activity, we calculate dynamic ranges for each, obtaining \textbf{B} (the homogeneous network) and \textbf{C} (the heterogeneous one).
    The heterogeneous HMN shows much broader distributions and much larger values of dynamic ranges. 
    These simulations were done on a network with $M_0=6$, $m=3$, $h=8$ (implying $N=M_0m^{h-1}=13122$), $\alpha=2$, $\phi=0.25$ (heterogeneous) and $\phi=0.5$ (homogeneous). These activities were taken from an average of $50$ trials, taking $T=10^4~\text{ms}$.
    }
    \label{figLRGS}
\end{figure}

 We first consider the more established scenario in which the entire network is externally driven.
 Following the standard definition~\cite{Kinouchi06a}, the dynamic range is quantified from the stimulus intensities corresponding to 10\% and 90\% of the response amplitude, denoted by $r_{0.1}$ and $r_{0.9}$, respectively (see Methods for details). Specifically, \[ \Delta = 10 \log \left( \frac{r_{0.9}}{r_{0.1}} \right), \] where $r_{0.1}$ and $r_{0.9}$ are obtained by interpolating the response curve (Fig.~\ref{figglobal_response_dyn}\textbf{A}). 

 For systems undergoing the conventional second-order phase transitions, the dynamic range exhibits a sharp maximum precisely at the critical point ($\lambda=1$, see black curve of Fig.~\ref{figglobal_response_dyn}\textbf{C}). The maximum values of dynamic range depend on the network dimensionality for a continuous spectrum of topological dimensions, revealing an increase in the maximum dynamic range as the topological dimension $D$ decreases (Fig.~\ref{figglobal_response_dyn}\textbf{D}). 

 This enhancement originates from systematic changes in the shape of the response curves. Networks with lower topological dimension exhibit a steeper increase in activity for weak stimuli, allowing them to distinguish subtle variations in input while delaying saturation at higher stimulus intensities (Fig.~\ref{figglobal_response_dyn}\textbf{B}). Consequently, these networks can achieve substantially larger dynamic ranges than their high-dimensional counterparts. 

In addition to the larger values of $\Delta$, low-dimensional networks exhibit a distinctive hallmark  of extended critical regimes, reminiscent of Griffiths-like phases. The dependence of $\Delta$ on the control parameter $\lambda$ exhibits broad, smooth maxima rather than the narrow peaks characteristic of conventional critical points (Fig.~\ref{figglobal_response_dyn}\textbf{C}). These properties indicate that the enhanced sensitivity associated with critical dynamics is no longer confined to a single finely tuned parameter value and persists over an extended region of parameter space.

 The  mechanism underlying these changes in the curves of dynamic range as a function of $\lambda$ as the dimensionality of the system decreases (Fig.~\ref{figglobal_response_dyn}\textbf{C}) is attributed to the heterogeneous activations imposed by HMN. In low-dimensional networks with broad  distributions of connectivity (in-degree) and convergence index $\kappa$ (Table~\ref{table2}), heterogeneity is meaningful, and the activity spreads locally through structurally heterogeneous pathways, so different regions become activated at different stimulus intensities (Figs.~\ref{figtime series+rasters} and~\ref{figmod_activity}). As the external input increases, regions and modules are recruited progressively rather than simultaneously, generating a broad distribution of local responses that enhances the global dynamic range. In contrast, higher-dimensional networks allow activity to spread more uniformly across the system (Figs.~\ref{figtime series+rasters} and~\ref{figmod_activity}), leading to a more homogeneous activation pattern and a narrower range of distinguishable inputs. Therefore, the locality of interactions and the dynamical heterogeneity characteristic of sparse HMNs naturally give rise to enhanced stimulus discrimination without requiring fine-tuning to a single critical point.

\subsection{Module-specific critical slowing down enhances dynamic range}

Mirroring the heterogeneity observed in endogenous activity (Figs.~\ref{figtime series+rasters} and~\ref{figmod_activity}), responses to external stimulation also vary considerably across modules. Figure~\ref{figLRGS} quantifies this variability by displaying the dynamic range of individual modules at each hierarchical level.  Variability in dynamic range increases toward more internal levels (higher $\ell$), although its magnitude depends strongly on network architecture. In this case, the dynamic range varies from the mean by at most 3\% ($<1$ dB) across modules in the homogeneous network, whereas this variation reaches up to 23\% (6 dB) in the heterogeneous network.

\begin{figure}
    \centering
    \includegraphics[width=0.8\linewidth]{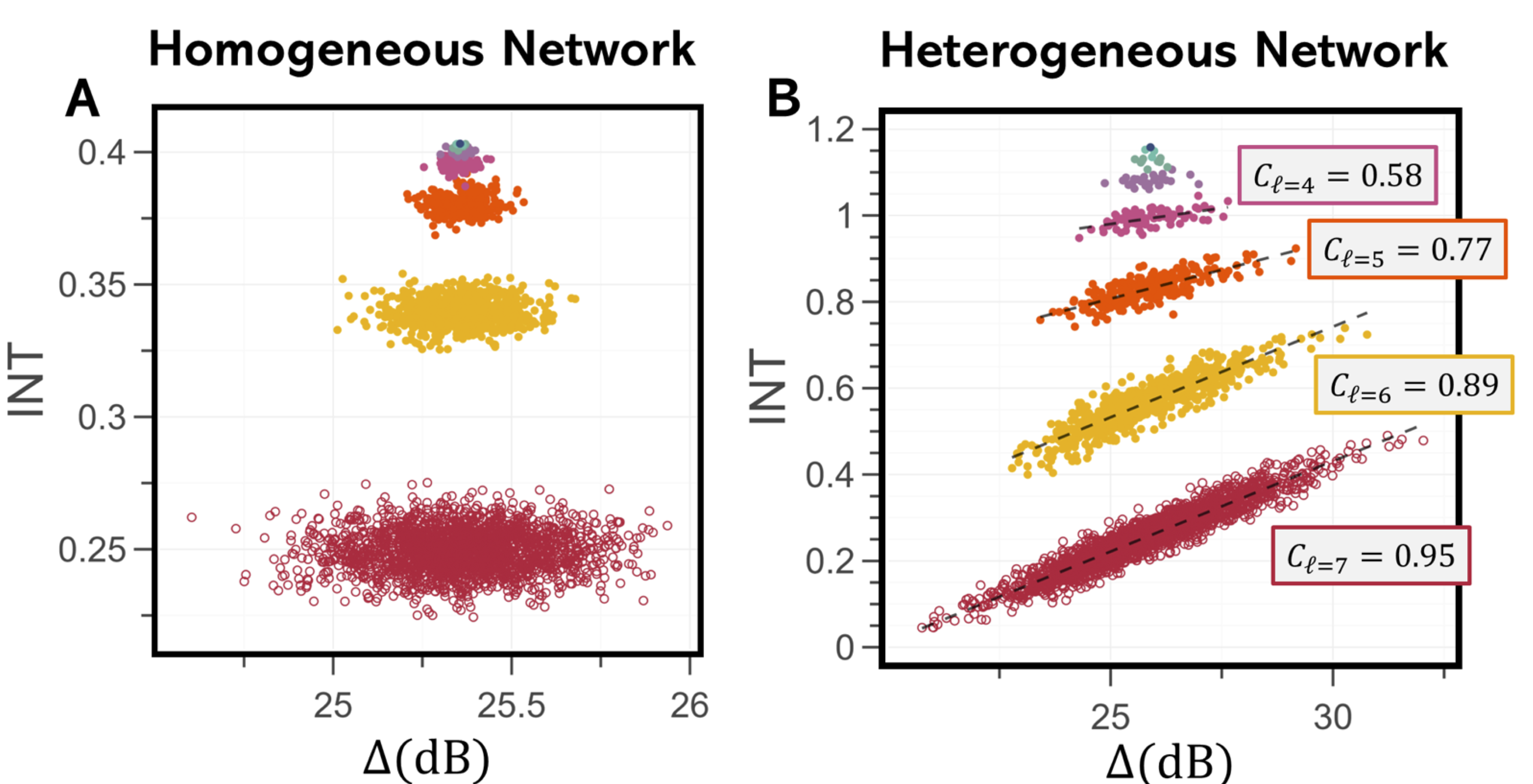}
    \caption{\textbf{Correlation between INT and dynamic range}. Homogeneous (left) and heterogeneous (right) networks under global stimulus. INT increases with hierarchical level  $\ell$ (indicated by different colors). Within a given hierarchical level (same color), only the heterogeneous network exhibits modules with positive correlations between INT and $\Delta_\ell$. 
    The correlations are stronger for lower hierarchical levels (deeper modules with larger values of $\ell$). 
    The networks shown are the same as those presented in  Figs.~\ref{figmod_activity} and~\ref{figLRGS}. 
    }
    \label{figDelta_vs_INT}
\end{figure}

This architecture-dependent heterogeneity extends to the relationship between intrinsic timescales and stimulus discrimination.
A positive correlation between INT and dynamic range emerges exclusively in heterogeneous networks (Fig.~\ref{figDelta_vs_INT}), suggesting that local criticality can induce module-specific critical slowing down. Modules operating near their local critical points exhibit prolonged intrinsic timescales, which enhance their dynamic range when the global network is tuned close to a transition~\cite{Kinouchi06a}. These results indicate how heterogeneity in the modular activity can diversify both temporal dynamics and functional responsiveness across modules.

\subsection
{Functional specialization and an optimal spatial extent for stimulus discrimination}

The results presented in Fig.~\ref{figLRGS} characterize the mean network activity of individual modules across hierarchical levels in response to stimulation of the entire network. Given the pronounced heterogeneity in the responses across modules, we next consider the complementary scenario in which stimulation is spatially localized. In the brain, external stimuli typically affect different neurons and regions to different degrees, reflecting functional segregation in which distinct regions are selectively sensitive to different types of stimuli. To reproduce this feature, we study localized stimulation, in which only specific modules at different hierarchical levels are stimulated. 

\begin{figure}
    \centering
    \includegraphics[width=0.6\textwidth]{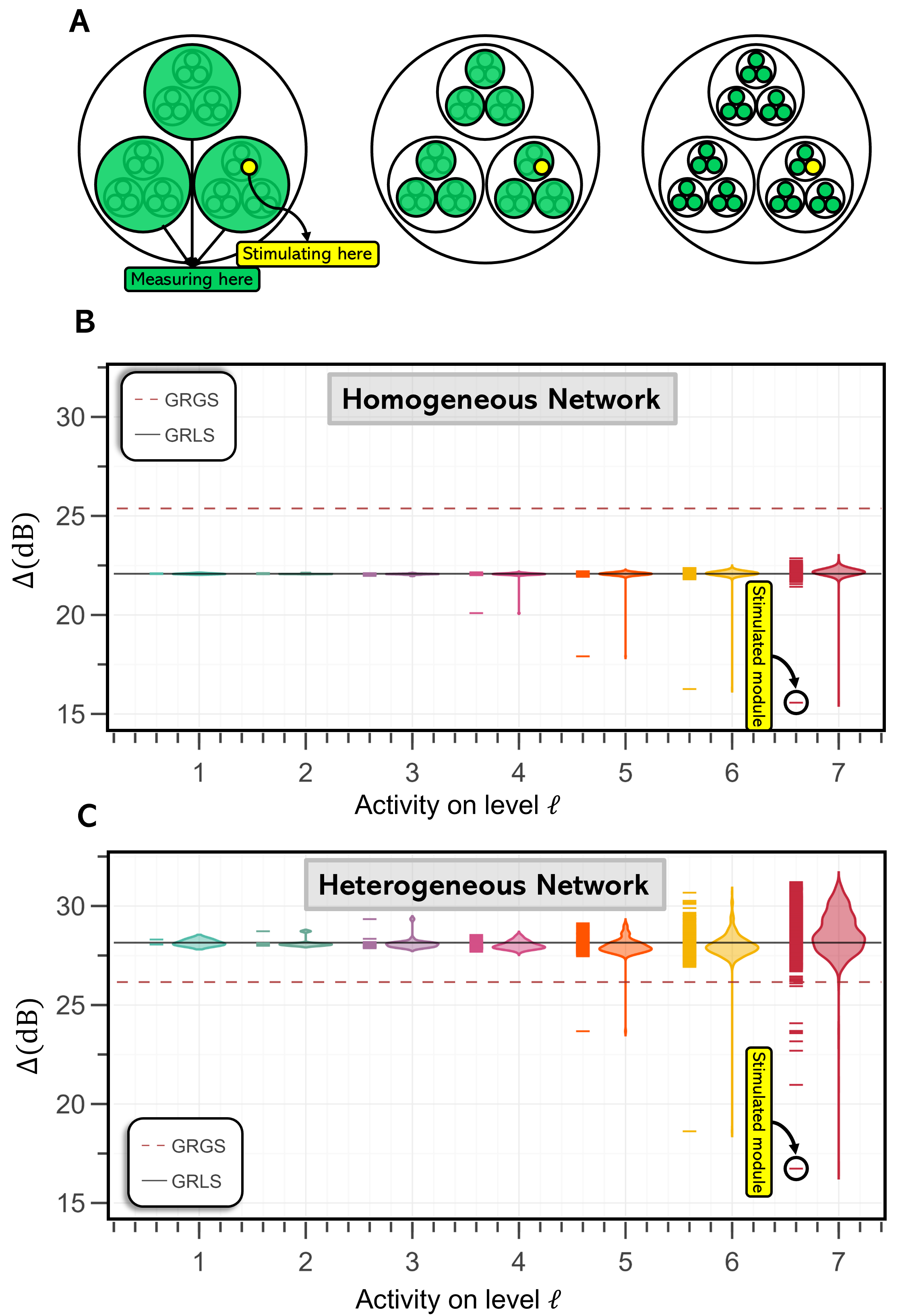}
    \caption{\textbf{Local response across modules and hierarchical levels to a stimulus selective for one innermost module.} Panel \textbf{A} Schematic representation of which modules are stimulated and from which level the measurement is taken, when the stimulation is always applied to the same module. By grouping the response curves as to the hierarchical level of activity, we calculate dynamic ranges for each group, obtaining \textbf{B} for a homogeneous network and \textbf{C} for a heterogeneous network. Both panels show the dynamic range of the whole network ($\ell=0$) while subjected to that local stimulus ($\ell'=h-1=7$), denoted GRLS (global response to local stimulus, solid black line), together with  GRGS, the dynamic range obtained through global stimulation (dashed red line). The heterogeneous network shows an increase in global dynamic range when the stimulus is local, whereas the homogeneous network shows a decrease in global dynamic range following local stimulation.  Simulations were performed on a network with $M_0=6$, $m=3$, $h=8$, $\alpha=2$, $\phi=0.25$ (heterogeneous) and $\phi=0.5$ (homogeneous). Activities were averaged over $50$ trials, taking $T=10^4~\text{ms}$.}
    \label{figLRLS}
\end{figure}

We first localize the stimulation to a single innermost module and measure activity at different hierarchical levels. 
Figure~\ref{figLRLS} illustrates this experiment and presents the resulting dynamic ranges corresponding to the response functions obtained for the different modules. 
This approach allows us to compare the Global Response to the Global Stimulus (GRGS), obtained by stimulating all nodes and averaging activity over the entire network, as shown in~\ref{figLRGS}, with the Global Response to a Local Stimulus (GRLS), defined as the activity of the entire network when only one of the innermost modules is stimulated. 
A key advantage of the heterogeneous network is that local stimulation produces a larger dynamic range than global stimulation, yielding $\Delta_{\mathrm{GRLS}} > \Delta_{\mathrm{GRGS}} $. This is shown by the  horizontal lines in Fig.~\ref{figLRLS}\textbf{B,C}, in which the red dashed line represents $ \Delta_{\mathrm{GRGS}} $ and the black line $\Delta_{\mathrm{GRLS}}$. Moreover, the dynamic range at the innermost level, $\ell=7$, has a larger variability and can attain much  larger values than at the outer levels (with smaller values of $\ell$). 

\begin{figure}
    \centering
    \includegraphics[width=0.6\textwidth]{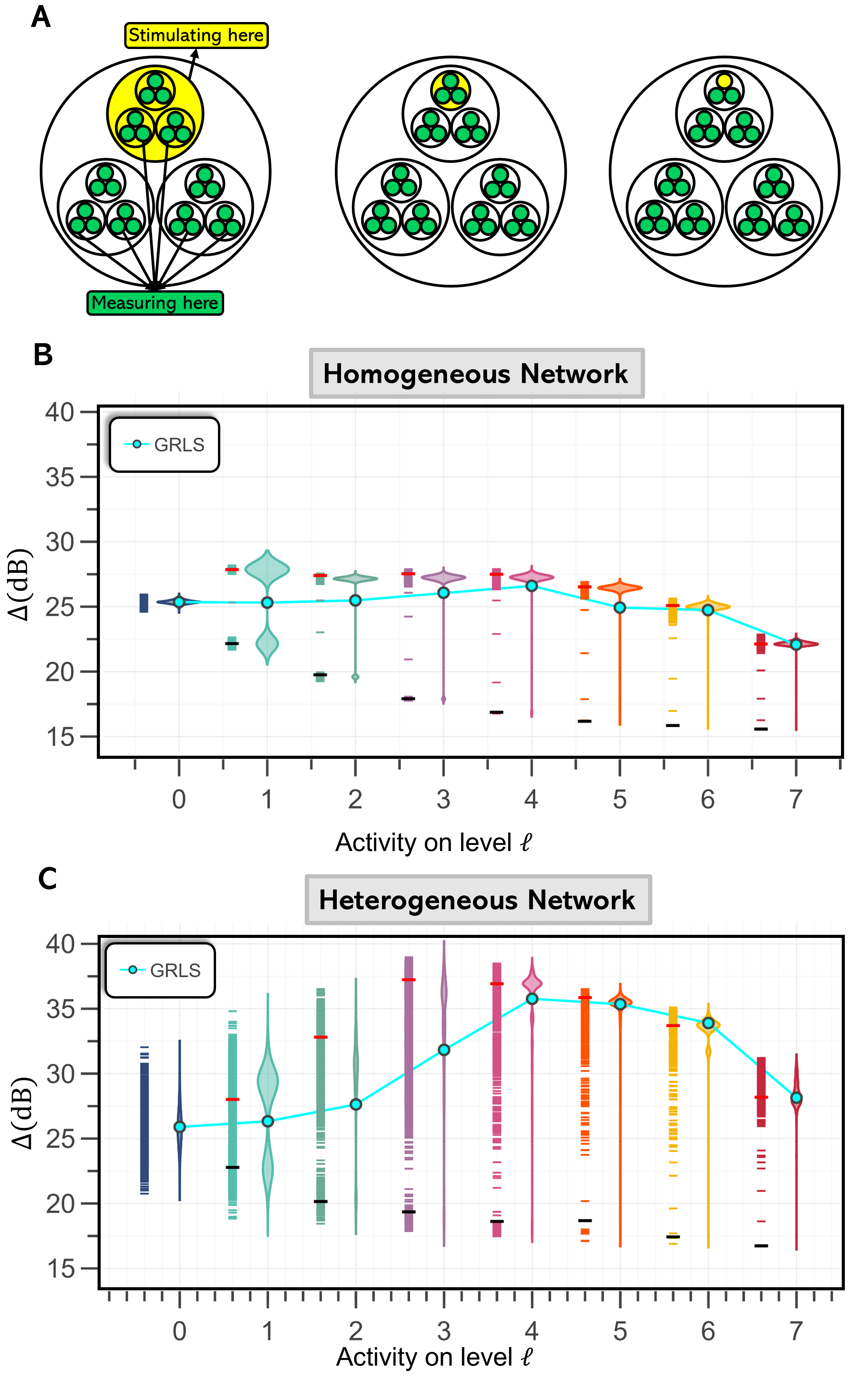}
    \caption{\textbf{Local response across modules to a stimulus at varying hierarchical levels.} \textbf{A}:Schematic representation of which modules are stimulated and from which level the measurement is taken, highlighting that activity is always sampled at the innermost modules. \textbf{B}, \textbf{C}: By grouping each response curve obtained, dynamic ranges are assessed, yielding \textbf{B} for the homogeneous network and \textbf{C} for the heterogeneous one. Both panels also show the dynamic range of the whole network ($\ell=0$) while subjected to that local stimulus ($\ell$), denoted by GRLS (global response to local stimulus). This reveals a peak in global dynamic range when the stimulus is at $\ell'=h/2=4$, in both homogeneous (\textbf{B}) and heterogeneous (\textbf{C}) HMNs, while the latter displays much more significant gains in $\Delta$, with larger values and broader distributions.  Simulations were performed on networks with $M_0=6$, $m=3$, $h=8$, $\alpha=2$, with $\phi=0.25$ for the heterogeneous network and $\phi=0.5$ for the homogeneous network. Activities were averaged over 50 trials, taking  $T=10^4~\text{ms}$.}
    \label{figLRVS}
\end{figure}

As described above, the protocol in Fig.~\ref{figLRLS} stimulates only a single innermost module, which corresponds to $M_0=6$ neurons. To assess how the spatial extent of the stimulated region affects network responsiveness, we exploit the hierarchical organization of the network. Specifically, we progressively vary the stimulation set, beginning with stimulation of all modules ($\ell'=0$) and restricting the stimulus to successively deeper hierarchical levels until only a single innermost module ($\ell'=7$) is stimulated. Throughout this protocol, we measure the response firing rate at all innermost modules, including the stimulated module. This allows us to compute the dynamic range and isolate how the spatial extent of the stimulus, rather than the extent of the measurement, controls the network's sensitivity, as illustrated in Fig.~\ref{figLRVS}. 

The results in Fig.~\ref{figLRVS} reveal a pronounced and non-monotonic dependence of the global dynamic range on the hierarchical level of stimulation. In particular, the curve of $\Delta_{\mathrm{GRLS}}$ exhibits a distinct peak when stimulation occurs at the intermediate level $\ell'=h/2=4$. This peak is accompanied by increasingly diverse and broadly distributed local dynamic ranges as stimulation is restricted to deeper hierarchical levels, as illustrated in Fig.~\ref{figLRVS}\textbf{B,C}. Although the peaks of the $\Delta_{\mathrm{GRLS}}$ curves coincide at $\ell'=4$, its mean value   is substantially higher in heterogeneous networks, which display both larger dynamic ranges and broader distributions across modules for all level of stimulation. 
These results indicate that maximal global responsiveness does not  arise from either global stimulation or the most localized stimulus, but  instead emerges from an intermediate spatial extent of the input to heterogeneous networks.

\begin{figure}
    \centering
    \includegraphics[width=1.0\linewidth]{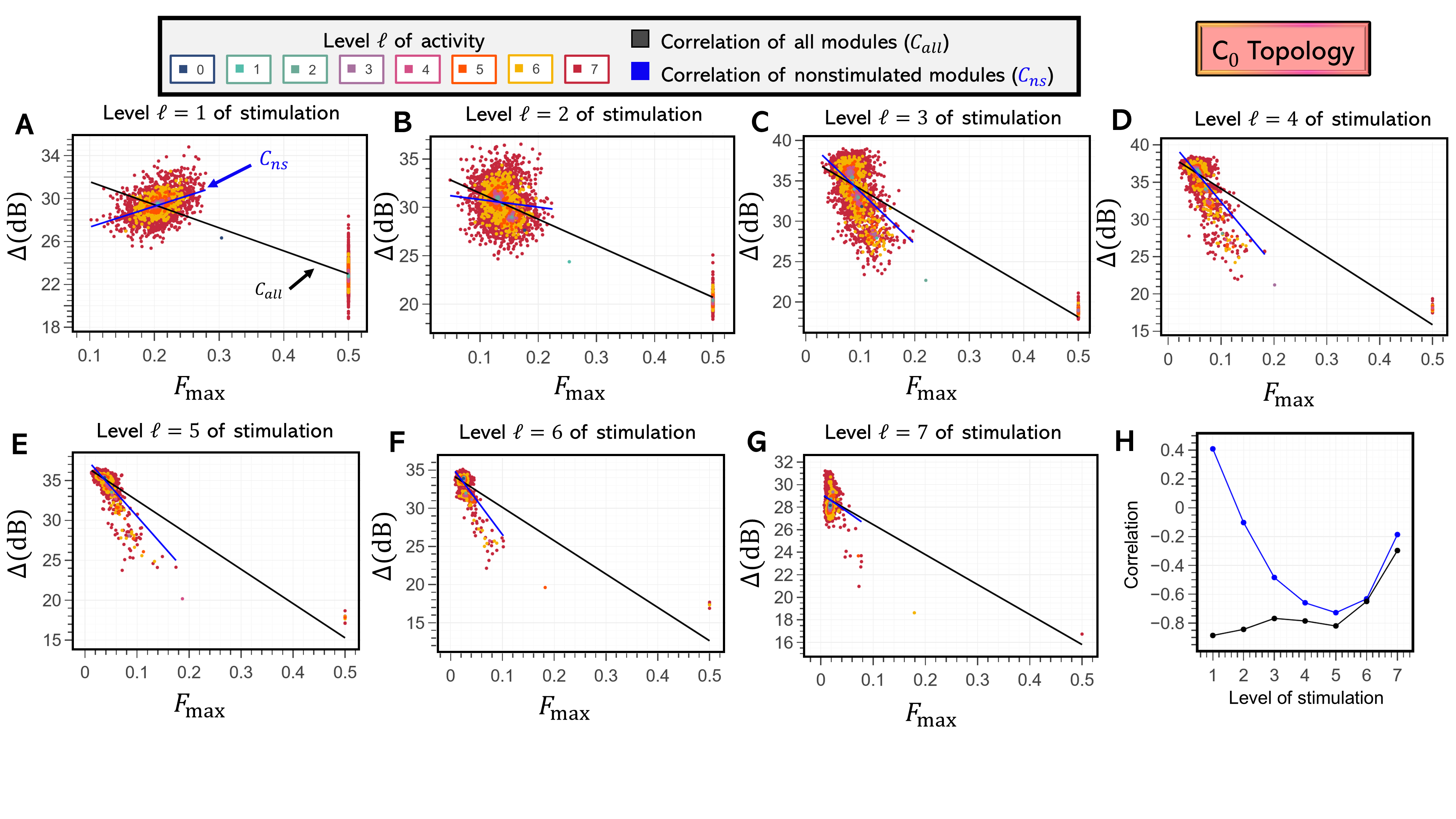}
    \caption{\textbf{Modular dynamic range vs. $F_{\max}$ for each module.} From \textbf{A} to \textbf{G}, what varies is the level $\ell$ of stimulation, similar to Fig.~\ref{figLRVS}. Both blue and black lines show pearson (linear) correlation sampled from the innermost activity between $\Delta$ and $F_{\max}$, whereas the blue line correlates only the non-stimulated modules and the black line correlates all modules. This shows that even by stimulating 1/3 of the network, $C_{all}$ is significant, and that aligns with gains on some local modules, as seen on Fig.~\ref{figLRVS}. Simultaneously, \textbf{H} attests that the minimum of $C_{ns}$ aligns with the peak of $\Delta _{GRLS}$ seen on Fig.~\ref{figLRVS}\textbf{C}, and displays the response curve role on increasing $\Delta$. Same data as in Fig.~\ref{figLRVS}.}
    \label{figDel_vs_F_max}
\end{figure}

A consistent consequence of local stimulation is an increase in $F_{\max}$ in the stimulated regions. 
This systematic association between $F_{\max}$ and $\Delta$ suggests that changes near the saturation of the response function play an important role in enhancing the dynamic range under local stimulation. Figure~\ref{figDel_vs_F_max} examines this relationship, showing that modules receiving external input exhibit larger $F_{\max}$ values and, consequently, smaller dynamic ranges.

This division between directly stimulated and non-stimulated modules gives rise to a natural form of functional specialization. Stimulated modules saturate early and respond best to weak stimuli, whereas unstimulated modules remain sensitive across a much broader range of intensities. Localized stimulation thus does more than redistribute activity: it partitions the network into distinct subpopulations. This may also explain why discrimination peaks at an intermediate stimulus extent ($\ell=h/2$). Global stimulation saturates all modules and collapses this division of labor, whereas stimulating too few modules leaves too little of the network in a sensitive regime to support a broad response. The peak therefore reflects a trade-off between engaging enough modules and leaving enough unstimulated to preserve sensitivity. The resulting functional specialization is thus not a fixed architectural property but a dynamical regime that emerges when a stimulus recruits a sufficient, but not excessive, fraction of the network.

\section{Discussion}

The architecture of the brain, characterized by modular and hierarchical organization alongside pronounced heterogeneity across spatial scales, plays a crucial role in flexible information processing. Using HMNs, our study demonstrates that heterogeneous dynamics can emerge from disordered connectivity within these networks, fundamentally reshaping the interplay between network structure, critical-like dynamics, and stimulus discrimination.

The emergence of this dynamical heterogeneity is governed by the sparsity parameter $\phi$: sparser HMNs (smaller $\phi$) exhibit stronger heterogeneous connectivity and dynamics, and it is precisely in this regime that the mean-field approximation breaks down. This effect is further amplified in lower-dimensional networks, where GP occurs. Notably, substantial heterogeneity persists even in the limit of infinite-dimensional networks, indicating that it is not merely a finite-size or low-dimensional artifact.

Rather than producing a single critical point at which the entire network exhibits maximal sensitivity, dynamical heterogeneity distributes distinct dynamical regimes across modules, generating broad critical-like dynamics, heterogeneous INT, and substantial variability in stimulus responsiveness. This local heterogeneity confers a functional advantage: modular dynamic range is enhanced in modules with longer INT, and, under localized stimulation, in modules that do not receive direct input. Furthermore, the spatial scale of stimulation strongly shapes network responsiveness, with stimulation at intermediate hierarchical levels yielding the largest global dynamic range. These findings indicate that hierarchical-modular organization provides a structural substrate for functional specialization, supporting flexible, spatially distributed information processing without requiring the network to be precisely tuned to a single critical point.

\subsection{Functional specialization: Heterogeneity distributes critical-like dynamics across the network}

A central result of this study is that dynamical heterogeneity transforms the conventional picture of critical dynamics from a global phenomenon into a spatially distributed one, exhibiting functional specialization. In homogeneous or effectively infinite-dimensional networks, critical slowing down and dynamic-range enhancement are concentrated around a well-defined transition. In contrast, hierarchical-modular networks can exhibit broad regions of enhanced fluctuations, long intrinsic timescales, and power-law activity over an extended range of coupling strengths. This behavior is consistent with a Griffiths-like regime~\cite{Moretti2013griffiths}, in which structurally distinct regions can approach dynamical transitions at different parameter values, allowing them to optimize performance at different tasks~\cite{gollo2017coexistence}.

Our modular analysis provides a direct view of this mechanism. At a given global coupling strength, individual modules can occupy markedly different dynamical states, with some remaining predominantly subcritical while others already exhibit self-sustained activity. These differences are accompanied by broad distributions of intrinsic neural timescales, particularly in the deepest modules. Thus, the broad critical-like regime observed at the network level is not simply a broadened version of a homogeneous transition. Instead, it emerges from the coexistence of multiple locally distinct dynamical regimes~\cite{gollo2016diversity,gollo2017coexistence}.

This observation is important for interpreting criticality in biological networks~\cite{Fontenele2019criticality,castro2024}. If different brain regions have different effective coupling strengths, connectivity patterns, or dynamical states, then a single global measure may necessarily average across regions that are not equally close to a transition~\cite{harris2024tracking,cambrainha2026hierarchical,morales2023quasiuniversal}. The resulting global signature of criticality can therefore reflect the collective contribution of multiple local dynamical regimes rather than a single network-wide critical point. In this sense, heterogeneity provides a natural mechanism for reconciling critical-like dynamics with the biological variability and structural disorder of real neural systems.

The relationship between local activity and intrinsic timescales further supports this interpretation. Modules with higher endogenous activity exhibit longer intrinsic timescales, and this association is substantially stronger in heterogeneous networks (in contrast to single neuron results~\cite{wu2024dendrites}), showing that each spatial scale might have its own governing rules~\cite{zeraati2026neural,gao2020neuronal,cavanagh2020diversity}. Critical slowing down therefore becomes a local property that varies across the hierarchy. This distributed organization may be particularly relevant to the hierarchy timescales observed across cortical systems~\cite{kiebel2008hierarchy,murray2014hierarchy,gollo2015dwelling}, where regions with different anatomical and functional roles exhibit distinct temporal dynamics, and regions (or modules) closer to criticality would exhibit longer INT owing to critical slowing down~\cite{harris2024tracking,cambrainha2026hierarchical}.
\\

\subsection{From global criticality to recruitment in modular dynamic range}

The conventional notion of dynamic range treats the network as a single computational unit~\cite{Kinouchi06a,shew2009neuronal,Assis08,ribeiro2008deterministic,copelli2007excitable, Larremore11a,publio2012dynamic, gollo2012signal,gollo2009active,furtado2006response, gollo2012statistical}. Our results suggest that this perspective can obscure an important consequence of hierarchical organization: different modules can have substantially different capacities to encode external inputs, similarly to the idiosyncratic signatures of dynamic range also observed across neurons~\cite{kirch2020spatially,gollo2013single}. We therefore introduced modular dynamic range as a local measure of stimulus responsiveness. Under global stimulation, heterogeneous networks exhibit substantially broader distributions of dynamic range than homogeneous networks, with variability becoming particularly pronounced at deeper hierarchical levels. 

This heterogeneity is not merely a variability in output. It is systematically related to the intrinsic dynamics of individual modules. A positive association between intrinsic timescale and dynamic range emerges in heterogeneous networks, linking prolonged temporal integration to enhanced stimulus discrimination. This relationship suggests that local critical slowing down may have a direct computational consequence: modules that retain activity over longer timescales can remain sensitive to a broader range of input intensities.

The global enhancement of dynamic range in lower-dimensional networks provides a complementary perspective. As dimensionality decreases, activation becomes increasingly localized and heterogeneous, causing different regions to be recruited at different stimulus intensities. Rather than the entire network responding simultaneously, activity progressively propagates through structurally distinct regions, broadening the range over which changes in stimulus intensity can be discriminated. This is a proof-of-concept for the recruitment theory~\cite{cleland1999}, and potentially also meaningful to explain recruitment differences and differences in heterogeneity of responses observes in autism spectrum disorder compared to the neurotypical population~\cite{wertheimer2024}. This finding suggests a testable prediction: that brain networks in autism spectrum disorder might exhibit lower effective dimensionality than in the neurotypical population, a hypothesis that warrants future investigation. 

This spatially heterogeneous recruitment mechanism provides an intuitive interpretation of the Griffiths-like enhancement of dynamic range. In a homogeneous system, increasing the stimulus primarily changes the overall magnitude of activity. In a heterogeneous hierarchical system, increasing the stimulus can additionally change which modules become active and how strongly they contribute. The response curve therefore reflects a progressive recruitment of multiple local dynamical regimes. Thus, heterogeneity effectively distributes the sensitivity of the network across stimulus intensities.

\subsection{Local stimulation reveals a distributed computational benefit}

A distinctive implication of our results emerges when external stimulation is spatially localized. In homogeneous networks, stimulating only a single module reduces global dynamic range relative to stimulating the entire network. In heterogeneous networks, however, the opposite occurs: local stimulation can produce a larger global dynamic range than global stimulation. This result indicates that heterogeneity can convert spatially restricted inputs into enhanced network-level sensitivity.


This enhancement is not restricted to the directly stimulated module, and modules that receive no external input can also exhibit enhanced modular dynamic range under localized stimulation. This finding suggests that the computational consequence of local stimulation is distributed across the network rather than being confined to the site of input. Such a property is particularly relevant for hierarchical neural systems, in which information entering through a restricted subset of regions can influence distributed processing across multiple spatial scales, whilst increasing network-level dynamic range.

This result provides a possible mechanistic interpretation of functional specialization as a results of diversity in dynamics and function~\cite{gollo2017coexistence}. Anatomical modules need not be intrinsically dedicated to a single computational function solely through differences in their local response properties. Instead, their position within a heterogeneous HMN can determine how strongly they respond to particular spatial patterns of input. Structural organization can therefore sculp functional specialization through differences in dynamical susceptibility.

\subsection{Intermediate-scale stimulation maximizes responsiveness}

A particularly striking result is the non-monotonic dependence of global dynamic range on the hierarchical level at which stimulation is applied. Neither global stimulation nor stimulation of the smallest possible module maximizes network responsiveness. Instead, the largest global dynamic range occurs when stimulation is applied at an intermediate hierarchical level, approximately halfway through the hierarchy.

This finding suggests that optimal information transmission in HMNs require a balance between  competing effects. Stimulation that is too spatially broad reduces the opportunity for differential recruitment because much of the network receives the input simultaneously. Conversely, stimulation that is too localized may fail to engage sufficiently many pathways to generate a strong distributed response. Intermediate-level stimulation provides a compromise: it is sufficiently localized to preserve spatial heterogeneity while encompassing enough of the network to engage multiple downstream modules.
The fact that an optimum stimulation profile emerges at an intermediate level in both homogeneous and heterogeneous networks, while the magnitude and variability of the enhancement are substantially larger in heterogeneous networks, is also informative. The hierarchical architecture itself establishes a preferred spatial scale for information propagation, whereas heterogeneity amplifies the functional consequences of this organization.

This intermediate optimum is likely reminiscent of a matching between the spatial scale of an input and the intrinsic organization of the network. Such a mechanism could provide a simple explanation for why different hierarchical levels may be particularly effective for different classes of stimuli. Inputs acting at distinct spatial scales could preferentially engage different portions of the hierarchy, thereby producing distinct and flexible global response profiles without requiring structural changes in the underlying network architecture.

\subsection{Implications for brain organization}

Although the model is deliberately abstract, our results point to several principles that may be relevant to dynamics in biological HMNs such as the brain: (i) heterogeneity need not be merely a source of noise that neural systems must overcome; instead, structural heterogeneity can itself generate a repertoire of dynamical states supporting flexible computation; (ii) hierarchical organization can convert this heterogeneity into an ordered hierarchy of temporal and response properties; and (iii) localized inputs can exploit this organization to produce distributed responses whose sensitivity depends on both the spatial scale and the hierarchical position of the input.

These principles provide a potential mechanistic bridge between anatomical hierarchy, intrinsic timescales, and functional specialization. Brain regions with different hierarchical positions are known to exhibit different temporal dynamics~\cite{kiebel2008hierarchy,gollo2015dwelling, murray2014hierarchy, raut2020hierarchical, hasson2015hierarchical, muller2023spatial,golesorkhi2021temporal, cambrainha2026hierarchical, harris2024tracking}, while sensory inputs are typically neither perfectly global nor restricted to single neurons. Our results suggest that these two features (hierarchical timescales and spatially structured stimulation) can interact through heterogeneous network dynamics to improve stimulus sensitivity.

More broadly, the results support a view of criticality in which the computational benefits traditionally associated with a finely tuned critical point can emerge from an extended and spatially distributed regime. Rather than requiring the entire system to operate at a unique optimal coupling, different modules can occupy different portions of a broad critical-like landscape. This organization provides robustness to parameter variations while simultaneously increasing functional diversity. In this sense, disorder and heterogeneity may provide not only tolerance to imperfect tuning but also an additional computational degree of freedom.

\subsection{Limitations and future directions}

While the present study offers a controlled framework for isolating how structural heterogeneity and hierarchical organization shape local dynamics and dynamic range, several extensions would help establish the generality of these findings. Future work should validate these principles in biologically detailed spiking models, which may require incorporating inhibition, adaptation, and synaptic kinetics, and should extend the analysis to heterogeneous weighted synapses, spatially embedded networks, and realistic connectomes (as illustrated in Fig.~\ref{figConnectomes}). 

\subsection{Conclusion}
Our findings indicate that heterogeneity in hierarchical-modular networks has implications extending well beyond simply broadening critical-like dynamics. It produces a spatially distributed array of dynamical states, establishes a gradient of intrinsic timescales, and enhances stimulus responsiveness at both global and modular levels. Notably, localized stimulation can enhance the global dynamic range in heterogeneous networks, with peak responsiveness occurring when stimulation is applied at an intermediate hierarchical level. These effects arise from the progressive and heterogeneous recruitment of network regions, thereby linking hierarchical-modular organization to stimulus-dependent computation, in which functional specialization emerges naturally. This suggests that the brain's dynamical heterogeneity not only aligns with efficient information processing but actively supports robustness and computational advantage. By distributing critical slowing down and stimulus sensitivity across modules, hierarchical organization can provide multiple dynamical regimes within a single network, allowing different spatial scales of input to engage distinct computational pathways. Consequently, rather than requiring precise tuning to a single critical point, complex brain networks may leverage heterogeneity itself to achieve robust, flexible, and functionally specialized computation.

\section{Methods}

\subsection{Construction of hierarchical-modular networks}

HMNs can be described uniquely by a $N\times N$ adjacency matrix $\textbf{A}$, where the row and column indexes represent the $N$ neurons, with a nonzero entry for every connection. The coupling strength between neurons is then encoded by a fixed parameter, $p$, for the whole network (see \ref{subsec:timedyn} for details), giving rise to the connectivity matrix $\textbf{C} = p \textbf{A}$. The matrix $\textbf{C}$, then, not only encodes topology, but also the dynamics of the network.

In order to simulate the hierarchical functionality and structural capabilities of the brain connectome, we use the model described initially by Moretti and Muñoz~\cite{Moretti2013griffiths} and Geza Ódor, et. al~\cite{odor2015griffiths}. In it, the network is described by its adjacency matrix $\textbf{A}$, presenting $m$ modules (the innermost each with $M_0$ nodes) and $h$ hierarchies. Connections depend only on the hierarchical level $\ell \in [0,h)$, such that each module is represented by a random network such that the connection density within is given by $\phi _\ell = \alpha \phi^{h-\ell}$, where $\alpha$ and $\phi$ are control variables. Higher values of $\ell$ represent smaller probabilities (since $\phi<1$), implying narrower regions (in other words, the structure is coarser).

The parameter $\phi$ controls network sparsity: lower values produce sparser networks, while higher values yield denser connectivity. A key measure of this connectivity is the network's dimension, $D$, as defined by Daqing et al.~\cite{daqing2011dimension}.  This dimension is determined by an iterative process similar to Dijkstra's algorithm~\cite{dijkstra2022note}: A random neuron is chosen as the seed, and its neighbors are counted. The process then proceeds outward, counting the neighbors of those neighbors, while ensuring that no neuron is counted more than once at each step. This continues for $r$ steps, yielding the total number of unique nodes reached, $N_r$.  The network dimension $D$ is then defined by the power-law relationship $N_r\propto r^D$ (valid for finite dimension networks), loosely interpreted as the dimension of the fractal-like hypersphere of dimension $D$, as the space occupied by the nodes scales accordingly~\cite{emmerich2013complex}. Square and cube lattices serve as key references for $D=2$ and $D=3$, respectively. By contrast, infinite-dimensional networks  exhibit much faster growth than the power law, whereas for $D \rightarrow 0$, $N_r$ rapidly saturates at a threshold beyond which no new neurons can be reached. Fig. \ref{fig:network} presents a comparison of these cases.

Only when $\phi$ is in the vicinity of $2^{-m}$, this model can generate networks with a finite topological dimension, when the heterogeneous nature of the network is heightened. In this study, whenever we compare these infinite-dimensional models (that is, when $\phi > 2^{-m}$) with their finite-dimensional counterparts, we ensure a fair comparison by using the same underlying structure (the same $M_0$, $m$, and $h$ values) but different values for $\phi$. Additionally, the parameter $\alpha$ can be any positive real number, and its role is to determine the specific value of the network's dimension while keeping it finite. It is informative to picture these five network parameters in two categories (Fig.~\ref{fig:network}): the geometric parameters ($M_0$, $m$, and $h$), which influence the network's modularity, and the connection parameters ($\phi$ and $\alpha$), which govern how the nodes connect. 

Notably, HMNs are unique to this type of study by virtue of their very nature: modularity creates islands of stability through topology, while the hierarchies ensure that these regions remain intertwined, implying distinct levels of connectedness among them. The relationship between the modules and the hierarchies can be controlled via the parameters $\phi$ and $\alpha$, making them key to navigating through degrees of heterogeneity in the network.



\subsection{Empirical connectomes}
The empirical connectomes were binary networks of human, mouse, and \textit{C. elegans}, identical to those used in~\cite{zalesky2016connectome}, with the \textit{C. elegans} connectome obtained from~\cite{varshney2011structural} and the mouse connectome obtained from~\cite{oh2014mesoscale,van2016comparative}. These networks vary substantially in scale: a node in the human connectome represents a large brain region, whereas a node in the \textit{C. elegans} connectome represents an individual neuron. Despite these differences in scale and resolution, modular organization and hierarchical core-periphery structure are recurring organizational principles across species~\cite{van2016comparative,puxeddu2024relation,kaiser2023connectomes}.

\subsection{Graph-theoretic properties}
\label{subsecGT}
Graph-theoretic measures were computed for all directed binary networks using the Brain Connectivity Toolbox~\cite{rubinov2010complex} in \texttt{MATLAB}. Network density was defined as the ratio of the number of edges present in the network to the total number of possible edges, i.e., the fraction of realized connections out of all pairwise connections that could exist. In-degree was defined, for each node, as the number of incoming edges to that node (i.e., the number of other nodes with a direct connection projecting onto it). Network efficiency was quantified by global efficiency~\cite{rubinov2010complex,zalesky2016connectome}, network clustering by the mean clustering coefficient across all nodes~\cite{fagiolo2007clustering,rubinov2010complex}, and modularity by the Q-index from a generalized Louvain community detection algorithm~\cite{rubinov2010complex}. The Q-index quantifies the extent to which a network is modular, with positive values indicating greater modular organization than expected by chance. Finally, the convergence index $\kappa$ was defined, for each node, as the product of its in-degree and  $p$, the probability of activity propagating along a connection to a neighboring node; this measure was used to characterize the extent to which heterogeneity in in-degree across nodes contributes to heterogeneity in mean activation (firing rate) across the network.

\subsection{Temporal dynamics}
\label{subsec:timedyn}

For the  simulation of neuronal behavior, we use the Kinouchi-Copelli model \cite{Kinouchi06a}, where each neuron can possess one of 2 states, i.e., $s_i =\{ 0,1\} $, cyclically traversed in a synchronous update, as in every cellular automaton. The state $s_i =0$ implies rest, $s_i =1$ is the excited state. A neuron can assume the value $s_i =1$ in two ways: due to an external stimulus, modeled as a Poisson process with rate $r$ (that is, a probability $\eta (r)=1 - e^{-r})$; or with probability $p$ (which can be interpreted as a coupling strength between neurons), due to an excited neighbor at the previous step. After $s_i =1$, the automaton returns to state $s_i =0$ at the next time step.

Defining $\rho _t$ as the density of active sites, i.e., the fraction of sites $i$ with input $s_i = 1$, we can obtain the average activity F of the network: $ F=\frac{1}{T}\sum _{t=0}^T \rho _t$. A common feature of the dynamics is the presence of a transient state before reaching some better-defined behavior in $F$; this transient is excluded from the analysis in all results shown here. When a network is stimulated via a Poisson process, the curve of $F$ as a function of the stimulus rate $r$ is a response function, and it presents a region of optimized information output, defined in this context by Kinouchi \& Copelli~\cite{Kinouchi06a} as the dynamic range of the network. This region corresponds to the range of external stimuli for which the resulting neuronal activity is distinguishable, excluding stimuli that are too low, where small changes fail to significantly alter the network's activity, and stimuli that are too high, where the network has already reached its saturation level. The dynamic range is thus the region lying between these two regimes, in which small changes in the external stimulus are actually meaningful for distinguishing the mean network activity $F$ activity as  $r$ varies. See Fig.~\ref{figglobal_response_dyn}\textbf{A} for a visual outline of this region.

When the correlations between neurons and its connections are small, there is an analytical critical point of such models, and it occurs when the largest eigenvalue of \textbf{C}, i.e., $\lambda = \lambda _c ^0 = 1$ \cite{Larremore11a}. However, the  Griffiths phase can stretch this value beyond $\lambda _c ^0$ to a new threshold for supercritical regime, labeled $\lambda _c ^G > \lambda _c ^0$ \cite{Moretti2013griffiths, vojta2006rare}, implying a range of $\lambda$ values where power laws apply not only to the $\rho (t)$ time series, but also to avalanches~\cite{rusch2025influence} and response curves.

Different realizations of the same network ensemble (generated by the same stochastic rules) can exhibit distinct patterns of neural activity. Although all networks are drawn from the same ensemble and share identical construction parameters, different realizations of the adjacency matrix lead to differences in both the intensity and the spatial distribution of neural activity. To disentangle structural heterogeneity from the stochastic variability of the dynamics, all results reported here are obtained using a single fixed network realization, averaging exclusively over different initial neuronal configurations and stochastic dynamical realizations.

\subsection{External stimulus and dynamic range}

For the estimated response curves, $80$ logarithmically spaced values of $r$ were generated, spanning the interval $r \in [r_0 = 10^{-7}, r_f = 10^2]$. Each system iteration was run for a total time $T = 10^4$ms, with an initial transient period of $\tilde{t} = 100$ms discarded prior to the computation of the mean value of $F$. To ensure statistical accuracy, $60$ samples were collected for each value of $r$, and the average of $F$ was computed from these samples.

This procedure yields a sigmoidal curve of $F$ (on a linear scale) as a function of $r$ (on a logarithmic scale), allowing the estimation of $F_0 \equiv F(r \rightarrow 0)$ using a nonlinear fit, which explicitly accounts for the smoothing of $F$ as $r \rightarrow 0$. In this case, $F$ is locally approximated by the functional form $F \approx a \cdot r^b + c$, and a nonlinear fit is performed over the interval $r \in [r_0, 10 \cdot r_0]$, yielding the coefficients $a$, $b$, and $c$, with $F_0 = c$. 

The nonlinear fitting approach proved to be the most faithful, as it accounts for both the fluctuations in $F$ and its power-law decay trend as $r \rightarrow 0$, particularly near the trivial critical point associated with second-order phase transitions.
Given both $F_\text{max}$ and $F_0$, we obtain $F_x = F_0 + x(F_\text{max} - F_0)$. The values of $r_{0.9}$ and $r_{0.1}$ are then obtained by interpolating the data points $(F, r)$ that bound $(F_{0.9}, r_{0.9})$ and $(F_{0.1}, r_{0.1})$, respectively.


\subsection{Modular coordinates}
\label{subsecCoordinates}


Due to the network's hierarchy, modules inherently contain sub-modules, forming hierarchical layers. This means that, using Fig. \ref{fig:network}\textbf{B} as an example, the set of connections in $X$ belongs to an outer module, which in turn belongs to another one, and so on. Specifically, these outer regions are represented with a trailing sequence of zeros in the modular coordinate set, as indicated by the modules $Y$ and $Z$ on Fig. \ref{fig:network}\textbf{B},\textbf{C}. 
This method thus provides a way to sample specific modules independently.

As an example, we can use the network of $m=3$, $h=4$, seen on Fig. \ref{fig:network}\textbf{B},\textbf{C}. The n-tuple $[1,1,1]$ is composed of 3 numbers, meaning that this module lies at $\ell=3$, the deepest one. This module is contained within $[1,1,0]$, that is in turn contained within $[1,0,0]$, finally within $[0,0,0]$, the 3-tuple that represents the whole network. Generalizing this process to all modularization structures yields the chosen configuration with $m=3$ modules and $h=8$ hierarchies, implying that an example of an innermost module on such a network is the n-tuple (or, in this case, 7-tuple) $[1,1,1,1,1,1,1]$.





\section*{Acknowledgments}


This work was made possible by the generous participation of the Systems and Computational Neuroscience group at UFPE. Maria Amanda Pérez de Oliveira provided motivation for some research conducted here, and Gustavo Cambrainha contributed valuable suggestions and general assistance with data visualization and the organization of the manuscript.

\nolinenumbers

%
%

\bibliography{filipe}

\end{document}